\documentclass[referee,usenatbib]{mn2e}

\usepackage{graphicx,natbib,rotating}

\title[Pc-scale Magnetic-Field Structures in HEAO-1 BL Lacs]
{Parsec-scale Magnetic-Field Structures in HEAO-1 BL Lacs}
\author[P. Kharb, D. Gabuzda and P. Shastri]
{P. Kharb$^{1, 3}$\thanks
{Major part of this work was completed at IIA, India
and JIVE, Holland. 
Present Email: pkharb@physics.purdue.edu}, 
D. Gabuzda$^{2}$ and 
P. Shastri$^{3}$\\ 
$^{1}$Department of Physics, Purdue University, West Lafayette, IN 47907, USA\\
$^{2}$University College Cork, Cork, Ireland\\
$^{3}$Indian Institute of Astrophysics (IIA), Bangalore 560034, India}
\begin{document}


\pagerange{\pageref{firstpage}--\pageref{lastpage}} \pubyear{2002}

\maketitle

\label{firstpage}

\begin{abstract}
We present very long baseline interferometry polarization images of 
an X-ray selected sample of BL~Lacertae objects belonging to the 
first High Energy Astronomy Observatory ({\it HEAO-1}) and the
$ROSAT$--Green Bank (RGB) surveys. These are primarily 
high-energy-peaked BL~Lacs (HBLs) and exhibit core-jet radio morphologies 
on pc-scales. They show moderately polarized jet components, similar to 
those of low-energy-peaked BL~Lacs (LBLs). 
The fractional polarization in the unresolved cores of the HBLs is, on 
average, lower than in the LBLs, while the fractional polarizations in the 
pc-scale jets of HBLs 
and LBLs are comparable. However a difference is observed in the
orientation of the inferred jet magnetic fields -- while LBL jets are
well-known to preferentially exhibit transverse magnetic fields, the 
HBL jets tend to display longitudinal magnetic fields. 
Although a `spine-sheath' jet velocity structure, along with larger
viewing angles for HBLs could produce the observed magnetic field
configuration, differences in other properties of LBLs and HBLs,
such as their total radio power,
cannot be fully reconciled with the different-angle scenario alone. 
Instead it appears that LBLs and HBLs differ
intrinsically, perhaps in the spin rates of their central black holes.
\end{abstract}


\section{Introduction}

BL~Lacertae objects are radio-loud active galatic nuclei (AGNs) 
characterized by a predominantly non-thermal, highly polarized 
continuum, that is variable in total intensity and polarization at all 
observed wavelengths, and exhibits weak or no optical emission lines 
\citep{Strittmatter72,SteinOdell76}. These extreme phenomena are 
understood to be a consequence of relativistic beaming in their nuclei 
\citep{BlandfordRees78,BlandfordKonigl79}. This picture has 
lead to Unification schemes \citep{UrryPadovani95} wherein the 
low-luminosity Fanaroff-Riley type-I radio galaxies 
\citep[FRI,][]{FanaroffRiley74} 
are believed to be the parent population of BL Lac objects,
based on similarities in orientation-independent properties such as 
extended radio emission, emission-line luminosity, host-galaxy type and 
environment \citep{Browne83,WardleMooreAngel84,Prestage88,UrryPadovani95}. 

Most BL~Lacs discovered before the late 1980's were identified based on radio 
data \citep[e.g.,][]{LeddenOdell85,Burbidge87}. However, space-based 
X-ray surveys were highly efficient in discovering new BL~Lac objects 
\citep[e.g.,][]{Giommi89,Stocke89}. 
The observed spectral energy distributions (SEDs) of the X-ray-selected and 
radio-selected BL~Lacs are systematically different. The majority of 
X-ray-selected BL Lacs show a synchrotron emission peak in the UV/soft X-ray 
regime making them `high-energy peaked BL Lacs' or HBLs, while most 
radio-selected BL~Lacs show synchrotron emission peaks in the near-IR/optical
making them `low-energy peaked BL~Lacs' or LBLs
\citep{GiommiPadovani94,PadovaniGiommi95}. 
The approximate division between LBLs and HBLs has been defined as
log~($f_x/f_r$)~$\sim-5.5$ \citep{Perlman94,Perlman96,Wurtz96} 
where $f_x$ and $f_r$ are
the 1keV X-ray and 5~GHz radio flux density, respectively.
HBLs are defined as BL~Lacs that have log~($f_x/f_r$)~$\ge-5.5$. 

LBLs are typically more core-dominated on kpc-scales 
\citep{PerlmanStocke93,Kollgaard96,RectorStocke00,Giroletti04a}, they display 
higher average optical 
polarizations and greater variability \citep{Schwartz89,Jannuzi94}, lower
starlight fraction \citep{Morris91}, and have 
more powerful radio lobes than HBLs \citep{Kollgaard92,LaurentMuehleisen93}. 
Based on nuclear trends, it appears that the LBLs are more ``extreme'' than 
HBLs, which in turn has lead to the suggestion that LBLs are oriented at 
smaller angles to the line of sight than HBLs \citep{Stocke85}.
This scenario is supported by population studies 
\citep{UrryPadovani90,UrryPadovani91} and has been explained in the framework 
of an accelerating-jet model, wherein the X-ray photons are 
emitted in the slower part of the jet and are therefore less beamed than 
the radio photons, which are emitted in the faster portion of the jet
\citep{Ghisellini89,UrryPadovaniStickel91}. However, the inadequacy of the 
orientation scenario to explain the systematic differences in 
the SED peaks of the HBLs and LBLs has given rise to the alternative 
scenario wherein the X-ray emitting regions of HBLs have higher electron
Lorentz factors and/or magnetic field strengths than LBLs \citep{Sambruna96}.
\citet{Georganopoulos98} have invoked a model that couples jet orientation and 
electron kinetic luminosity to unify the BL~Lac subclasses.

The apparent ``blazar sequence'' of increasing synchrotron peak 
frequency with decreasing source luminosity, as observed in the average SEDs
of flat-spectrum radio quasars (FSRQs), LBLs and HBLs
\citep[see][]{Sambruna96,Fossati98,Maraschi01} has been challenged by the
discovery of high-energy peaked FSRQs \citep{Sambruna97,Perlman98,Padovani02}
and high luminosity HBLs \citep{Giommi05}. In addition,
the gap between LBLs and HBLs has now been filled with the discovery
of intermediate-peak BL~Lacs (IBLs) which have SED peaks in the optical/UV
wavebands and exhibit intermediate properties between LBLs and HBLs 
\citep{NassBade96,Kock96,Perlman98,LaurentMuehleisen99,Stevens99}. 
Thus the question of HBLs and LBLs being two
distinct AGN classes, as opposed to being the extreme ends of a 
continuous distribution of BL~Lac properties, is an open one.

Owing to their relatively greater radio flux densities,
LBLs have been the subject of extensive VLBI observations 
\citep[e.g.,][]{Gabuzda00,Kellermann04,Lister05}.
These studies have found them to be typically compact, with a core and a 
relatively small number of fairly discrete jet components. Measured 
superluminal speeds are typically modest, in the range of 2--4$c$  
\citep{Gabuzda94,Gabuzda00,Kellermann04}. VLBI polarization observations 
have shown a number of systematic differences between LBLs and FSRQs,
initially detected in images obtained at 5--8~GHz 
\citep{Gabuzda92,Cawthorne93,Gabuzda00} but generally being 
retained in the higher-frequency 15~GHz images \citep{Lister05} $-$
LBLs have appreciably polarized cores (core fractional polarization, 
$m_c \sim 2 - 5\%$) with predominantly transverse magnetic ($B$) fields in 
their jets, while quasars have weakly polarized cores ($m_c < 2\%$) and 
predominantly longitudinal $B$ fields in their pc-scale jets. 

HBL polarization properties have, on the other hand, not been explored as 
much with VLBI experiments due to their relatively lower radio flux densities.
5~GHz total intensity VLBI images of a sample of HBLs belonging to the 
$Einstein$ ``Slew" Survey were presented by \citet{Rector03}. They found that
HBLs displayed core-jet morphologies similar to the LBLs. However there was
a weak evidence that the HBL jets exhibited smaller misalignments between
parsec and kiloparsec scales than LBL jets \citep[see also][]{Giroletti04a}. 
This was interpreted as being 
either due to intrinsically straighter jets in HBLs or their jets
being oriented further from the line of sight than LBLs jets. 
Note however that 8.4~GHz VLA observations of a sample of EMSS and 
{\it HEAO-1} BL~Lacs failed to show a systematic trend of smaller
misalignments in HBL jets (E. Perlman, private communication).

In this paper, we present VLBI polarization observations of a sample of 
18 BL~Lacs belonging to the {\it HEAO-1} hard X-ray and the 
$ROSAT$--Green Bank (RGB) surveys. Thirteen of these BL~lacs have been 
classified as HBLs. Early VLBI polarization results for four of these 
{\it HEAO-1} BL~Lacs were presented by \citet{Kollgaard96}. 
\S~2 and 3 describe the BL~Lac sample and the observations.
We present our results in \S~4--6. A discussion follows in 
\S~7, and the conclusions in \S~8. We assume throughout a 
cosmology where $H_0$ = 71~km~s$^{-1}$~Mpc$^{-1}$,
$\Omega_M$ = 0.24, and $\Omega_{\Lambda}$ = 0.76. 

\section{The Sample}

The {\it HEAO-1} Large Area Sky Survey (LASS) 
detected 842 hard X-ray (0.8 $-$ 20 keV) sources over the entire sky
\citep{Wood84}. To date, 29 BL~Lacertae objects have been identified
in the LASS catalog \citep{LaurentMuehleisen93},
one of which (2201+044) is now known to be a Seyfert 1 galaxy 
\citep{Veron-Cetty93}. Note that 2201+044 is part of our sample.
The RGB sample of BL~Lacs was generated 
from a cross-correlation of the $ROSAT$ All-Sky Survey (RASS) and a 
reanalysis of the 1987 Green Bank 6-cm radio survey 
\citep{Gregory96,LaurentMuehleisen97}. 
VLA observations of the {\it HEAO-1} BL Lac sample have been 
presented by \citet{LaurentMuehleisen93} and \citet{Kollgaard96}.
Here we present VLBI polarization results for nine {\it HEAO-1} BL~Lacs, 
along with eight BL~Lacs belonging to the RGB sample.
In all, there are 13 HBLs, 4 LBLs and 1 Seyfert-1 galaxy.
Of the 4 LBLs, 1147+245 also belongs to the 1-Jy sample. 

The BL~Lac sample is presented in Table~\ref{sampleBL}, which has the following
columns: Cols.~(1) \& (2) IAU and other names, (3) BL~Lac classification 
based on the SED peak frequency and/or the 
log~($f_x/f_r$) criterion (see $cf.~\S$~1), 
(4) sample membership, with H denoting sources belonging to the {\it HEAO-1} sample, 
R to the RGB sample, and J to the 1-Jy sample, (5) membership of other
``heritage'' X-ray samples: RX = {\it ROSAT} All-sky Survey, 
1E, 2E \& ES = {\it Einstein}, second Image Proportional Counter (IPC) 
{\it Einstein} 
X-ray Survey, and {\it Einstein} ``Slew'' Survey, respectively, 
(6) redshift, (7) projected linear 
scale corresponding to an angular scale of 1~mas, (8) $-$ (10) the epochs 
for which results are presented here, and (11) \& (12) the references for the 
redshifts and BL~Lac classification, respectively. 

\section{Observations and Data reduction}

The polarization observations were made at 5~GHz with (i) a global VLBI array
on February 23, 1993 (1993.15) and (ii) on July 13, 1995 (1995.53)
and June 28, 1998 (1998.49) with the 10-element Very Large Baseline Array 
(VLBA)\footnote{The VLBA is operated by the National Radio Astronomy 
Observatory, which is a facility of the National Science Foundation operated 
under cooperative agreement by Associated Universities, Inc.}.
In all cases, each source was observed for a total of roughly 1.3 to 
2.8 hours, in 12--26 scans
distributed throughout the time the source was observable with all or nearly
all of the antennas in the VLB arrays.  Two intermediate frequencies (IF, or
equivalently, baseband converters) were recorded in each polarization, with 
8~MHz per IF, and a total aggregate bit rate of 128~Mbits/s. 
The data were correlated by the VLBA correlator in Soccoro. The preliminary 
calibration, fringe fitting, 
polarization calibration and imaging were done using the Astronomical Image 
Processing System (AIPS) following standard methods.  

\subsection{Global VLBI}

The global VLBI array used for the February 23, 1993 observations included
the Effelsberg (EB), Green Bank (GB), and Medicina (MC) telescopes, the 
phased Very Large Array (Y27), and the Hancock (HN), North Liberty (NL), 
Brewster (BR) and Owens Valley (OV) VLBA antennas. Effelsberg
was used as the reference antenna at all stages of the calibration.
The unpolarized source OQ~208 was used as the instrumental polarization 
($D$-term) calibrator in the AIPS task {\sc LPCAL}. The absolute 
electric-vector polarization angle (EVPA or $\chi$) calibration was
performed by comparing the total VLBI-scale and (simultaneously 
measured) VLA core polarizations for the compact polarized source OJ~287.  

\subsection{VLBA}

The July 13, 1995 and June 28, 1998 observations were obtained using the ten
telescopes of the American VLBA.  Los Alamos 
was used as the reference antenna at all stages of the calibration. 
The unpolarized source 3C84 and the nearly unresolved polarized source
0749+540 were used for the $D$-term calibration for the July 1995 and
June 1998 observations, respectively. 
Short 2--3-minute VLA snapshots of 10
{\it HEAO-1} HBLs, including five of the four HBLs for which 1995.53 
images are presented here, were made at 4.9, 8.4, and 15.0~GHz on
July 18, 1995, only a few days after the July 1995 VLBA observations. 
We used the results of these 4.9~GHz observations for OJ287 for the EVPA 
calibration of the VLBA data, assuming that the polarization position 
angle of OJ287 did not vary between the VLA and VLBI observations. 

Unfortunately, we did not have integrated polarization measurements
of any compact polarized sources observed during our June 1998 VLBA
run nearby in time to those VLBA observations. Instead, we applied the
EVPA calibration determined for VLBA observations in March and April 1998,
based on integrated polarization measurements within a few days of those
experiments and using the same reference antenna (Los Alamos). The EVPA
calibration corresponds to the phase difference between the right-circularly
polarized and left-circularly polarized signals at the reference antenna,
and has been shown to be constant to within a few degrees over as long as
several years, if no adjustments are made to the receiver affecting this
instrumental parameter \citep{ReynoldsCawthorneGabuzda01}. Our observations
were only a few months after those of \citet{Charlot06}; moreover, the EVPA
calibration for 5~GHz VLBI observations obtained in June 2000, based on
simultaneous VLA polarization measurements and also reduced using Los 
Alamos as the reference antenna, indicates that this calibration had
remained the same to within a few degrees (Gabuzda, O'Sullivan \& Gurvits, 
private communication), thus justifying the application of the 5~GHz EVPA 
calibration for the March--April 1998 VLBA observations to our 5~GHz data. 

\begin{table}
\caption{Properties of the sample BL~Lacs.}
\begin{center}
\begin{tabular}{cclcccccccccc}
\hline\hline
IAU  &Other&Class& Sample& X-ray& $z$ & Scale &\multicolumn{3}{c}{Epochs} &$z$ & Class \\
name &name &     &       & Sample&&(pc/mas)& 1993.15 & 1995.53 & 1998.49 & Ref & Ref \\\hline
0414+009&        &HBL &H &RX,2E& 0.287  &4.285 & X & X & & 1 & 15 \\
0652+426&4C~42.22&HBL &R &RX   & 0.0590 &1.127 & & & X & 2   & 16 \\
0706+592&        &HBL &H &RX   & 0.1250 &2.215 & X & X & & 2 & 16 \\
0749+540&4C~54.15&LBL &R &RX     &$>$0.20 &3.268 & & & X & 3   & 16 \\
0829+046& OJ049  &LBL &R &RX     & 0.1736 &2.918 & X & X & X &4& 16 \\
0925+504&        &LBL*&R &RX   & 0.3703 &5.090 & & & X & 5   & 17 \\
1011+496&        &HBL &H &ES    & 0.212  &3.268& & & X & 6    & 16 \\
1101+384&Mrk 421 &HBL &H &RX,ES& 0.0300 &0.593 & & X & X & 7 & 15 \\
1133+704&Mrk 180 &HBL &H &RX,ES& 0.0452 &0.877 & & & X  & 8  & 15 \\
1147+245&        &LBL &J & ...  &$>$0.20 &3.268 & & X & &9   & 16 \\
1215+303&ON~325  &HBL &H &ES    & 0.1300 &2.291 & X & & X & 10& 15 \\
1227+255&        &HBL*&R &RX   & 0.1350 &2.366 & & & X & 11  & 17 \\
1235+632&        &HBL*&H &RX,1E& 0.2969 &4.389 & X & &  & 10 & 18 \\
1553+113&        &HBL*&R &RX,ES& 0.3600 &4.998 & & & X & 12  & 15 \\
1727+502&I Zw 187&HBL &H &RX,ES& 0.0554 &1.062 & & X & X & 13& 16 \\
1741+196&        &HBL &R &RX,ES& 0.0840 &1.559 & & & X & 14  & 15 \\
1743+398&        &HBL &R &RX    & 0.2670 &4.068 & & & X & 2   & 16 \\
2201+044&4C~04.77&Sy-1&H &RX    & 0.0270 &0.535 & & & X & 1   & 19 \\ \hline
\label{sampleBL}
\end{tabular}
\end{center}
{
Notes $-$ *BL~Lacs which have been alternately classified as 
IBLs by \citet{Nieppola06}.
Last two columns list references for redshift and BL~Lac classification, respectively.
1- \citet{FalomoScarpa94}, 2- \citet{LaurentMuehleisen98},
3- \citet{StickelKuhr93}, 4- \citet{Falomo91}, 5- \citet{Henstock97},
6- \citet{Albert07}, 7- \citet{UlrichKinman75}, 8- \citet{Ulrich78},
9- \citet{StickelFried93}, 10-\citet{BadeBeckmann98},
11- \citet{NassBade96}, 12- \citet{HewittBurbidge89}, 13- \citet{Oke78},
14- \citet{HeidtNilsson99}, 15- \citet{Donato05}, 16- \citet{Nieppola06}, 
17- \citet{LaurentMuehleisen99}, 18- \citet{Wurtz96}, 19- \citet{Veron-Cetty93}.}
\end{table}

\begin{table}
\begin{center}
\caption{Details of the VLBP observing program.}
\begin{tabular}{lllll}\hline\hline
Observing   &Freq     & Antennas&$D$-term&EVPA\\
date       & MHz      &         &calibrator&calibrator\\\hline
23 Feb 1993 & 4979.99 & MC, EB, WB, GB, Y27&OQ~208&0235+164,OJ~287\\
           &          & HN, NL, BR, OV&&\\
13 July 1995& 4987.49    & VLBA         &3C~84&OJ~287\\
28 June 1998& 4991.46    & VLBA         &0749+540& Other\\
\hline
\label{tab:logBL}
\end{tabular}
\end{center}
\end{table}

\section{Results}

\begin{figure*}
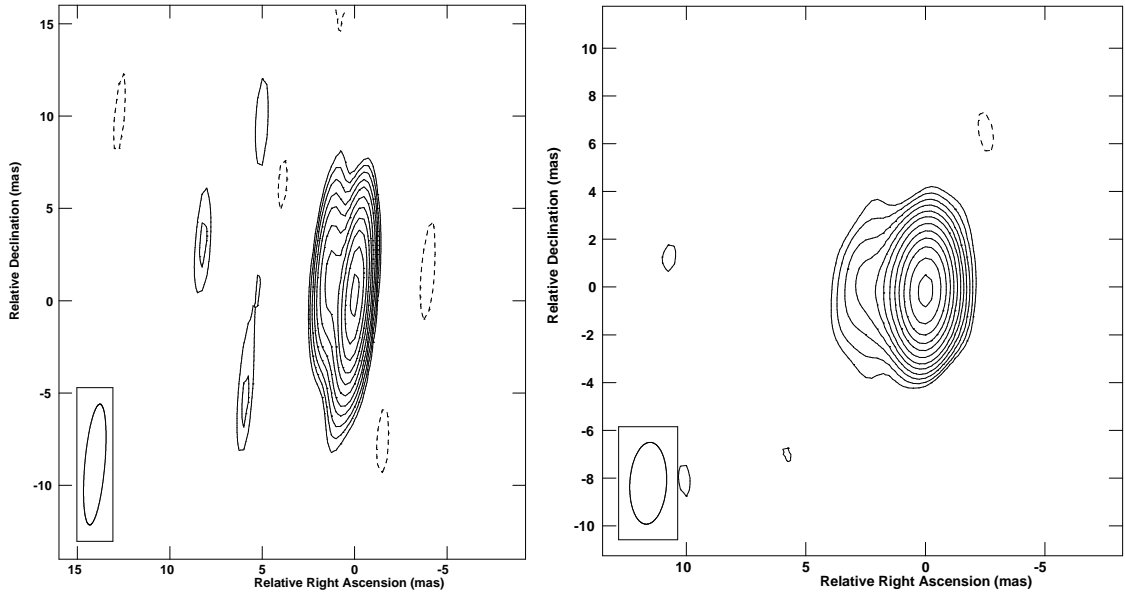

\centering{
\includegraphics[height=8.0cm]{gk7-0414-I.ps}
\includegraphics[height=8.0cm]{bk33-0414-I.ps}}
\caption{Total intensity image of the HBL 0414+009.
(Left) Epoch 1993.15.  Contours are --2.8, 2.8, 5.6, 11.2, 22.5, 45 and 
90\% of the peak surface brightness of 27.2 mJy beam$^{-1}$.
(Right) Epoch 1995.53. 
Contours are --2, 2, 2.8, 4, 5.6, 8, 11, 16, 23, 32, 45, 64, 90\%
of the peak surface brightness of 38.6 mJy beam$^{-1}$.}
\label{fig:0414}
\end{figure*}

The VLBI polarization (VLBP) observations of the 17 sample BL~Lacs (and 1 Sy-1 
galaxy) reveal core-jet morphologies in all but two sources, viz., 0706+592
and 0749+540. Polarization is detected in the cores and/or jets of all but 
three sources, viz., 0414+009, 0706+592 and 1235+632. Total intensity 
images with polarization electric vectors superimposed are presented in
Figs.~\ref{fig:0414} to \ref{fig:2201}. The Gaussian restoring beam for each 
object is shown in the lower left-hand corner of the image. 
The rms noise in the total intensity maps is typically $\sim 100~\mu$Jy/beam. 

The polarization electric vectors have not been corrected for Faraday rotation,
occurring either in our Galaxy (i.e., an overall rotation of the EVPAs
applied to all regions in the source) or in the immediate vicinity of the 
BL~Lac object (likely affecting only certain regions in the source).  
Multi-frequency VLBP observations have demonstrated that the contribution 
from regions of thermal magnetized plasma in the immediate vicinity of the 
BL~Lac object may be substantial in the core region, but is likely to be 
relatively small in the jet at appreciable distances from the core 
\citep[e.g.,][]{ZavalaTaylor03,ZavalaTaylor04}. Typical Galactic (i.e.,
foreground) rotation measures are no more than a few dozen rad/m$^2$,
corresponding to rotations at 5~GHz of about 10$^\circ$ or less 
\citep[e.g.,][]{Pushkarev01}. Thus,
we expect that the jet EVPAs should not be far from their intrinsic
values, while the core EVPAs may in principle be subject to appreciable
Faraday rotation. 

Models for all the sources were derived by fitting the complex, fully
calibrated total
intensity ($I$, the $RR$ and $LL$ correlations) and linear polarization 
($P$, the $RL$ and $LR$ correlations) visibilities that result from the mapping 
process with a small
number of circular Gaussian components, as described by \citet{Roberts87}
and \citet{GabuzdaWardle89}. We checked for consistency between the model fits,
the distribution of CLEAN components, and the visual appearance of the images.
The $I$ and $P$ visibilities were fitted separately, in order to allow for 
small differences in the positions and sizes of corresponding $I$ and $P$
components, either intrinsic to the source structure or associated with
residual calibration errors. 
The model-fitting results are presented in Table~\ref{modelBLL}, where the
columns list the fitted parameters for various epochs. Column (1) presents
the source name, while the remaining columns present for each component
(2) an identifying label, with C denoting the core; (3)--(4) the total
($I$) and polarized ($p$) intensities; (5)--(6) the polarization position angle
($\chi$) and fractional polarization ($m$); (7)--(8) the separation
from the core ($r$) and its uncertainty ($\Delta r$); (9)--(10) the structural
position angle ($\theta$) and its uncertainty ($\Delta\theta$); and the size
(FWHM).

Comments on the individual sources follow in \S~\ref{sec:notes}. 
When interpreting the observed polarization images,
we assume that the jet emission is optically thin; in contrast, the observed 
``core'' emission may be predominantly optically thick or optically thin, 
depending on the relative contributions of the optically thick ``intrinsic'' 
core and regions of the optically thin innermost jet that are blended with 
this feature due to our finite spatial resolution. 

\begin{table}
\caption{{\it HEAO-1} BL~Lac VLBI models}
\begin{center}
\begin{tabular}{lccccccccccccccc}
\hline\hline
Source &  Comp&$I$&$p$&$\chi$&$m$&$r$&$\Delta r$&$\theta$&$\Delta\theta$& FWHM\\
   &     &(mJy)&(mJy)&(deg)&(per cent)&(mas)&(mas)&(deg)&(deg)&(mas)\\
\hline 
0414+009 & \multicolumn{9}{c}{{\em 1993.15}}\\
&C&26.9&...&...&...&...  &...    &... &...&  0.02\\
&K1&8.3&...&...&...&1.43 &  0.06 & 72 &  5&  0.96\\
&\multicolumn{9}{c}{{\em 1995.53}}\\
&C&38.5&...&...&...&...  & ...  & ... & ...& 0.22\\
&K1&7.7&...&...&...&1.74 & 0.23 &  74 &  8 & 2.02\\ \hline
0652+426 & \multicolumn{9}{c}{{\em 1998.49}}\\
&C & 132.7 &...  &... &...  &...  &...   & ...   &...  &  0.09\\
&K5& 18.7  &0.71&-3.2 &3.79 &1.55 & 0.03 &   39  &   1 &  0.18\\
&K4& 4.7   &... &... &...   &3.08 & 0.12 &   28  &   2 &  0.07\\
&K3& 4.4   &... &... &...   &6.09 & 0.30 &   30  &   2 &  1.45\\
&K2& 2.5   &... &... &...   &5.99 & 0.20 &   35  &   2 &  0.29\\
&K1& 2.9   &... &... &...   &11.69& 1.84 &   28  &   6 &  4.27\\ \hline
0706+592 &\multicolumn{9}{c}{{\em 1993.15}}\\
&C &24.9&...&...&...&...    &...    &...       & ...    & 0.23\\
&\multicolumn{9}{c}{{\em 1995.53}}\\
&C &27.7&...&...&...&...   &...   &...   & ...   &0.19\\
&K1&1.9 &...&...&...& 4.05 & 0.50 & -157 &  7    &1.00\\ \hline
0749+540 &\multicolumn{9}{c}{{\em 1998.49}}\\
&C &1530.1&148.5&62.7&9.7&...  &...   &...   & ...  &0.14\\
&K3&93.2  &...  &... &...&0.75 & 0.03 & -124 & 3    &0.10\\
&K2&82.6  &...  &... &...&0.85 & 0.03 &  22  & 4    &0.29\\
&K1&7.8   &...  &... &...&4.31 & 0.26 &   5  & 6    &2.12\\ \hline
0829+046 &\multicolumn{9}{c}{{\em 1993.15}}\\
&C &470.7&15.2&24.9   &3.2  &...  &...       &...     &...&0.39\\
&K4& 80.8&3.2 &4.9    &4.0  &1.00 &   0.03   & 55 & 2 &0.39\\
&K3&189.6&11.1&--20.8 &5.8  &2.62 &   0.01   &  59 & 0 &1.22\\
&K2& 32.9&15.6&--65.9 &47.4 &4.08 &   0.06   &  50 & 1 &1.65\\
&K1& 20.6& ...&...    &...  &5.21 &  0.10    &  51  & 1 &2.74\\
&\multicolumn{9}{c}{{\em 1995.53}}\\
&C &674.9&20.1&55.1&2.9     &...  &...   &...  &...   & 0.45\\
&K5&161.2&9.9 &6.9 &6.1     &1.06 & 0.01 & 58  & 1    &0.79\\
&K4&150.2&8.9 &--21.5&5.9   &2.58 & 0.01 & 66  & 0    &0.44\\
&K3&48.9 &9.8 &--40.1&20.0  &3.97 & 0.02 & 57  & 0    &0.31\\
&K2&18.2 &5.6 &--39.5&30.8  &5.28 & 0.07 & 60  & 1    &0.37\\
&K1&14.3 &1.6 &1.2&11.1     &7.12 & 0.08 & 59  & 1    &1.10\\
&\multicolumn{9}{c}{{\em 1998.49}}\\
&C &414.0 &18.2&36.2&4.4    &...   &...    &...   &...   & 0.22\\
&K6&98.3  &10.7&--4.7&10.9  &1.02  & 0.02  &  57  & 1   &0.48\\
&K5&160.7 &7.9 &--47.5&4.9  &2.52  & 0.01  &  67  & 0   &1.03\\
&K4&24.1  &5.9 &--70.7&24.5 &3.90  & 0.04  &  61  & 1   &0.48\\
&K3&15.2  &...&...&...      &5.41  & 0.04  &  62  & 1   &0.41\\
&K2&17.7  &7.7&--42.6&43.5  &7.37  & 0.13  &  58  & 1   &2.35\\
&K1&9.9   &...&...&...      &9.88  & 0.19  &  58  & 2   &1.68\\
\hline
0925+504 &\multicolumn{9}{c}{{\em 1998.49}}\\
&C &477.1&67.8&58.7&14.2&...  &...   &...     &...& 0.57\\
&K2&19.1 &... &...&...  &6.47 & 0.18 &  128   & 1 &3.32\\
&K1&9.8  &... &...&...  &9.41 & 2.07 &  132   & 11&4.53\\
\hline
1011+496 &\multicolumn{9}{c}{{\em 1998.49}}\\
&C &89.3&4.2&--12.2&4.7  &...  &...   &...   &... &0.36\\
&K4&25.9&0.9&--10.3&3.5  &1.00 & 0.01 & -110 &  1 &0.27\\
&K3&9.3 &...&...&...     &2.12 & 0.03 & -96  &  1 &0.23\\
&K2&2.1 &...&...&...     &3.27 & 0.13 & -87  &  5 &0.06\\
&K1&8.2 &0.5&41.9&6.1    &6.62 & 0.09 & -102 &  1 &1.86\\
\hline
\end{tabular}
\end{center}
\end{table}
\addtocounter{table}{-1}
\begin{table}
\begin{center}
\caption{{\it HEAO-1} BL~Lac VLBI models (contd)}
\begin{tabular}{ccccccccccc}
\hline\hline
Source & Model&$I$&$p$&$\chi$&$m$&$r$&$\Delta r$&$\theta$&$\Delta\theta$& FWHM\\
   &  &(mJy)&(mJy)&(deg)&(per cent)&(mas)&(mas)&(deg)&(deg)&(mas)\\
\hline 
1101+384 &\multicolumn{9}{c}{{\em 1995.53}}\\
&C &318.4&4.8&--87.8&1.5   &...   &...    &...   &... & 0.17\\
&K8& 35.2&2.7&21.8  &7.7   &1.24  &  0.02 & -33  & 1  &0.17\\
&K7& 12.1&...&...   &...   &2.81  &  0.11 & -36  & 2  &1.43\\
&K6&  9.2&2.5&--43.2&27.2  &5.37  &  0.11 & -37  & 1  &1.49\\
&K5&  8.2&...&...&...      &9.95  &  0.76 & -32  & 4  &4.65\\
&K4&  6.9&...&...&...      &15.95 &  0.78 & -34  & 3  &4.43\\
&K3&  6.7&...&...&...      &19.19 &  0.55 & -47  & 2  &3.66\\
&K2&  1.6&0.7&--52.5&43.8  &20.52 &  0.63 & -62  & 2  &1.55\\
&K1& 26.1&...&...&...      &32.13 &  1.08 & -62  & 2  &14.9\\
&\multicolumn{9}{c}{{\em 1998.49}}\\
&C &348.3&4.2&--54.3&1.2    &...   &...    &...  &... & 0.19\\
&K8&31.8 &4.9&--2.6 &15.4   &1.35  & 0.03  & -40 & 1  & 0.43\\
&K7&5.1  &...&...   &...    &2.96  & 0.15  & -43 & 3  & 0.76\\
&K6&18.3 &1.8&--71.0&9.8    &5.44  & 0.14  & -41 & 1  & 2.41\\
&K5&18.9 &...&...   &...    &11.96 & 1.41  & -37 & 5  & 9.48\\
&K3&4.1  &1.1&--88.0&26.8   &19.63 & 0.61  & -52 & 2  & 3.35\\ \hline
1133+704 &\multicolumn{9}{c}{{\em 1998.49}}\\
&C &97.5&...&...&...      &...   &...    &...   &...& 0.21\\
&K6& 9.8&2.8&3.9&28.6     &1.01  &  0.10 & 103  & 4 & 1.13\\
&K5& 7.6&1.1&--39.6&14.5  &2.99  &  0.13 & 106  & 1 & 0.04\\
&K4& 2.1&...&...&...      &4.72  &  0.51 & 92   & 3 & 0.66\\
&K3& 4.3&...&...&...      &10.54 &  1.54 & 78   & 7 & 6.11\\
&K2& 1.6&...&...&...      &16.89 &  1.24 & 77   & 4 & 2.68\\
&K1& 0.4&...&...&...      &24.77 &  1.12 & 68   & 2 & 0.03\\ \hline
1147+245 &\multicolumn{9}{c}{{\em 1995.53}}\\
&C &432.1&18.6&10.9&4.3     &...   &...   &...    &...& 0.34\\
&K6&120.9&... &...&...      &0.88  & 0.01 & -103  & 1 & 0.66\\
&K5& 55.1&2.9 &--84.1&5.3   &2.65  & 0.02 & -96   & 1 & 1.61\\
&K4& 21.1&2.2 &--60.0&10.4  &8.35  & 0.32 & -90   & 1 & 4.20\\
&K3& 7.0 &4.1 &--59.4&58.6  &8.50  & 0.11 & -88   & 1 & 1.37\\
&K2&24.4 &2.9 &18.7&11.9    &13.73 & 0.12 & -100  & 0 & 3.45\\
&K1&21.8 &2.2 &20.6&10.1    &20.03 & 0.22 & -106  & 1 & 5.22\\ \hline
1215+303 &\multicolumn{9}{c}{{\em 1993.15}}\\
&C&34.2&...&...&...&...   &...     &... & ... & 0.29\\
&K5&3.6&...&...&...&0.84  &  0.07  & 94 & 13  &0.39\\
&K4&2.0&...&...&...&2.34  &  0.39  & 105& 30  &0.67\\
&\multicolumn{9}{c}{{\em 1998.49}}\\
&C &224.2&...&...&...     &...   &...    &...   &... & 0.21\\
&K5&30.9 &6.8&29.5&22.0   &1.39  &  0.02 &  146 & 1  & 0.42\\
&K4&5.0  &...&...&...     &3.18  &  0.13 &  143 & 2  & 0.56\\
&K3&5.4  &...&...&...     &6.09  &  0.22 &  139 & 2  & 1.55\\
&K2&2.0  &...&...&...     &10.04 &  0.17 &  138 & 1  & 0.07\\
&K1&7.2  &...&...&...     &15.48 &  0.50 &  143 & 2  & 3.32\\
\hline
1227+255 & \multicolumn{9}{c}{{\em 1998.49}}\\
&C &180.7&4.6&28.2  &2.5   &...   &...   &...       &... & 0.12\\
&K4&14.8 &3.7&30.8  &22.0  &1.14  & 0.03 & $-$127 & 2  &0.23\\
&K3&9.9  &1.5&3.1   &10.6  &2.35  & 0.05 & $-$117 & 2  &0.88\\
&K2&15.8 &1.7&--76.2&22.4  &4.32  & 0.17 & $-$114 & 2  &3.71\\
&K1&15.8 &1.2&87.3  &9.7   &7.87  & 0.16 & $-$127 & 1  &4.09\\
\hline
1235+632 & \multicolumn{9}{c}{{\em 1993.15}}\\
&C&14.1&...&...&...&...   &...   & ... &... & 0.16\\
&K1&1.9&...&...&...&1.74  & 0.12 & 150 & 5  &0.19\\
\hline
1553+113 & \multicolumn{9}{c}{{\em 1998.49}}\\
&C &182.9&4.2&22.8&2.3 &...   &...    &...  &...& 0.24\\
&K4&19.0 &2.8&3.3&14.7 &0.68  &  0.03 & 53  & 3 &1.07\\
&K3&8.9  &...&...&...  &3.24  &  0.24 & 20  & 4 &3.82\\
&K2&4.6  &...&...&...  &10.02 &  0.92 & 58  & 7 &6.22\\
&K1&2.4  &...&...&...  &97.03 &  1.20 & 85  & 1 &5.85\\
\hline
\end{tabular}
\end{center}
\end{table}
\addtocounter{table}{-1}
\begin{table}
\begin{center}
\caption{{\it HEAO-1} BL~Lac VLBI models (contd)}
\begin{tabular}{ccccccccccc}
\hline\hline
Source & Model&$I$&$p$&$\chi$&$m$&$r$&$\Delta r$&$\theta$&$\Delta\theta$& FWHM\\
   &  &(mJy)&(mJy)&(deg)&(per cent)&(mas)&(mas)&(deg)&(deg)&(mas)\\
\hline 
1727+502 &\multicolumn{9}{c}{{\em 1995.53}}\\
&C &76.2&1.2&--1.6&1.6 &...  &...   &...    &... & 0.28\\
&K3&26.6&1.4&44.5&5.3  &1.78 & 0.03 & --45  & 1  &0.96\\
&K2&17.1&1.5&27.3&8.8  &3.25 & 0.06 & --34  & 1  &1.59\\
&K1&27.2&1.0&62.1&3.7  &5.89 & 0.22 & --49  & 2  &4.08\\ 
&\multicolumn{9}{c}{{\em 1998.49}}\\
&C &59.0&...&...&...     &...  &...   &...   & ...&0.37\\
&K5&29.3&0.6&--60.3&2.0  &1.15 & 0.01 & --80 & 1  &0.91\\
&K4&19.9&2.1&3.8&10.6    &2.74 & 0.03 & --57 & 1  &1.08\\
&K3&19.8&2.3&42.9&11.6   &4.67 & 0.03 & --62 & 1  &1.43\\
&K2&16.1&...&...&...     &6.74 & 0.07 & --56 & 1  &2.48\\
&K1&15.6&...&...&...     &15.74& 0.27 & --51 & 1  &6.11\\ \hline
1741+196 &\multicolumn{9}{c}{{\em 1998.49}}\\
&C &84.2&0.9&--55.0&1.1  &...   &...   &...  &... & 0.38\\
&K6&15.9&...&...&...     &1.14  & 0.03 & 73  & 2  &0.53\\
&K5&14.4&1.4&--33.7&9.7  &2.75  & 0.03 & 79  & 1  &1.11\\
&K4&7.3 &0.8&--29.9&10.9 &5.87  & 0.11 & 84  & 2  &2.23\\
&K3&2.1 &...&...&...     &9.37  & 0.56 & 84  & 4  &2.78\\
&K2&1.6 &...&...&...     &13.39 & 1.79 & 88  & 7  &4.37\\
&K1&4.5 &...&...&...     &19.53 & 1.25 & 95  & 4  &7.39\\ \hline
1743+398 &\multicolumn{9}{c}{{\em 1998.49}}\\
&C &47.7&0.7&54.6&1.5   &...  &...    &...    &... & 0.06\\
&K2&7.0 &...&...&...    &1.60 &  0.07 & --179 & 1  &0.48\\
&K1&3.7 &1.0&28.6&27.0  &5.25 &  0.25 &  178  & 1  &1.48\\ \hline
2201+044 &\multicolumn{9}{c}{{\em 1998.49}}\\
&C &155.4&...&...&...     &...  &...    &...   &... & 0.39\\
&K2& 31.0&0.76&--50.9&2.5 &1.19 &  0.04 & --43 &  2 &0.21\\
&K1&  8.2&...&...&...     &3.39 &  0.08 & --48 &  1 &0.74\\
\hline
\label{modelBLL}
\end{tabular}
\end{center}
\end{table}

\section{Notes on individual BL~Lacs}
\label{sec:notes}

\subsection{0414+009} 
This BL~Lac object is hosted by a luminous elliptical that is the dominant
member of a galaxy cluster \citep{FalomoTanzi91}. The 5~GHz VLA image of
\citet{Reid99} shows a jet-like feature at position angle PA $\sim40\degr$.
Both of our VLBI images (Fig.~\ref{fig:0414}) show a jet in PA$\sim75\degr$, 
with no detected polarization.

\begin{figure*}
\centering{
\includegraphics[height=8cm]{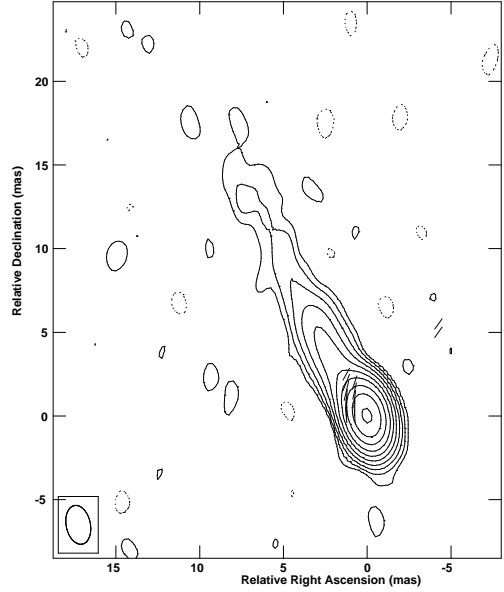}}
\caption{\small Total-intensity image of the HBL 0652+426, epoch 1998.49, 
with $\chi$ vectors superimposed.
Contours are --0.17, 0.17, 0.35, 0.70, 1.40, 2.80, 5.60, 11.20, 22.50, 45
and 90\% of the peak surface brightness of 136.6 mJy beam$^{-1}$,
$\chi$ vectors: 1 mas = 1 mJy beam$^{-1}$.}
\label{fig:0652}
\end{figure*}

\subsection{0652+426} 
The 1.4~GHz VLA image of this source shows a two-sided jet source embedded in a
bright halo, with the brighter jet at PA=40$\degr$ \citep{Rector03}. 
Our VLBI image (Fig.~\ref{fig:0652}) shows a well-collimated 
jet extending from the core
in PA$\sim40\degr$. Polarization is detected in the inner jet, with the
inferred $B$ field showing no obvious relation to the jet direction. 

\begin{figure*}
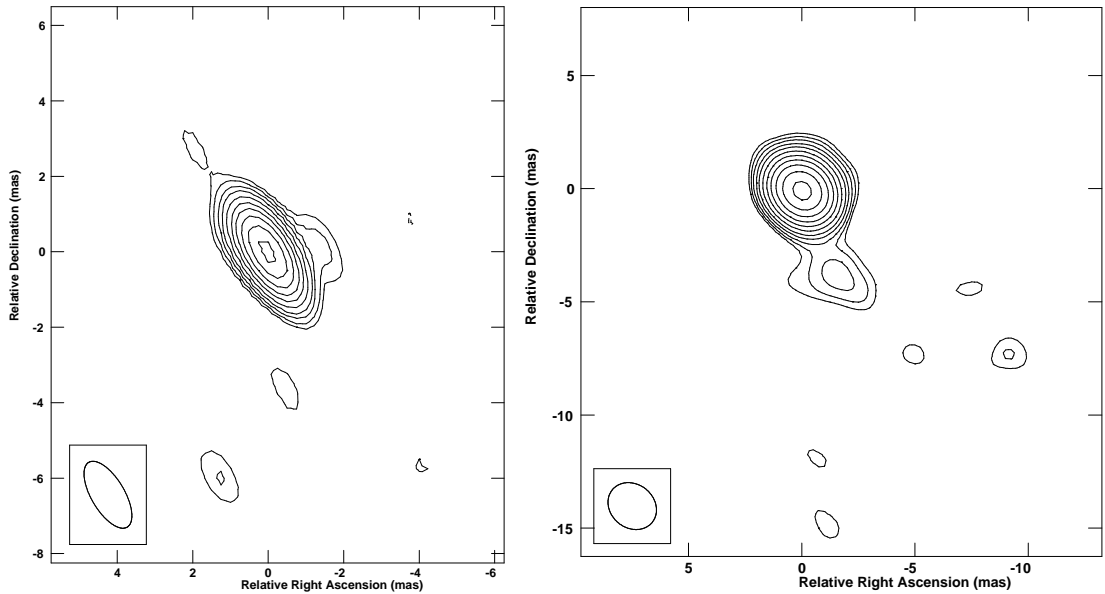

\centering{
\includegraphics[height=8.0cm]{gk7-0706-I.ps}
\includegraphics[height=8.0cm]{bk33-0706-I.ps}}
\caption{
Total intensity images of the HBL 0706+592. (Left) Epoch 1993.15.
Contours are --4, 4, 5.6, 8, 11, 16, 23, 32, 45, 64, 90\% of the peak surface
brightness of 23.9 mJy beam$^{-1}$.
(Right) Epoch 1995.53.  
Contours are --2.8, 2.8, 4, 5.6, 8, 11, 16, 23, 32, 45, 64, 90\%
of the peak surface brightness of 27.5 mJy beam$^{-1}$.}
\label{fig:0706}
\end{figure*}

\subsection{0706+592} 
The 1.4~GHz VLA image of \citet{Giroletti04a} shows the radio core to be
located on the northwest edge of a roughly spherical cocoon. They suggest 
that this object could be a wide-angle or narrow-angle tailed object, 
viewed at a small angle. Their 5~GHz VLBA image shows a weak jet to the
southwest. Our two VLBI images (Fig.~\ref{fig:0706}) are dominated by a 
compact core, with weak extentions to the west (in 1993) and southwest 
(in 1995); given the image of \citet{Giroletti04a}, the former is probably 
an artefact. We did not detect polarization at either of our epochs.

\begin{figure*}
\centerline{
\includegraphics[height=8cm]{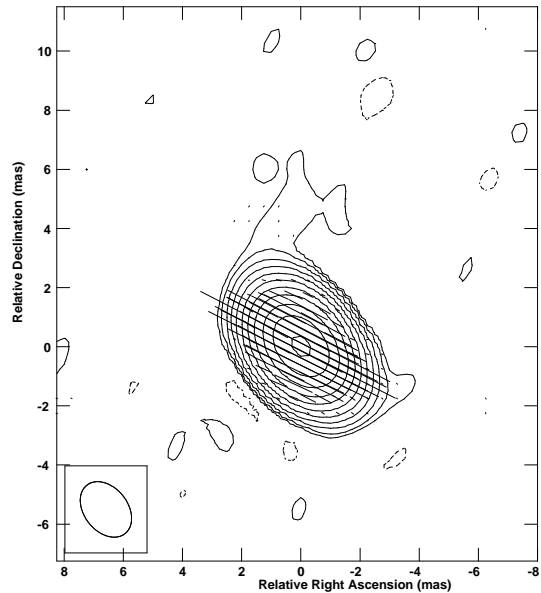}}
\caption{\small Total-intensity image of the LBL 0749+540, epoch 1998.49,
with $\chi$ vectors superimposed.
Contours are --0.09, 0.09 0.17, 0.35, 0.70, 1.40, 2.80, 5.60, 11.20, 22.50, 45
and 90\% of the peak surface brightness of 163.3 mJy beam$^{-1}$,
$\chi$ vectors: 1 mas = 20 mJy beam$^{-1}$.}
\label{fig:0749}
\end{figure*}

\subsection{0749+540} 
This source was classified as a BL~Lac object by \citet{KuehrSchmidt90} 
based on its featureless optical spectrum and high optical polarization.
A lower limit of $z>0.2$ for its redshift was derived by
\citet{StickelKuhr93} based on the absence of any extended structure of a host
galaxy in a direct image taken with the 3.5m Calar Alto telescope. 
The 5~GHz VLBI image of \citet{Taylor94} shows a jet emerging from the
core in PA$\sim0\degr$.  Our VLBI image (Fig.~\ref{fig:0749})
reveals a dominant core, with 
an extension to the north, in roughly the direction of the previously
detected VLBI jet. The core shows an unusually high fractional polarization 
of $\sim 9\%$, while no polarization was detected in the jet. 

\begin{figure*}
\centering{
\includegraphics[height=8.0cm]{gk7-0829-IP.ps}
\includegraphics[height=8.0cm]{bk33-0829-IP.ps}
\includegraphics[height=8.0cm]{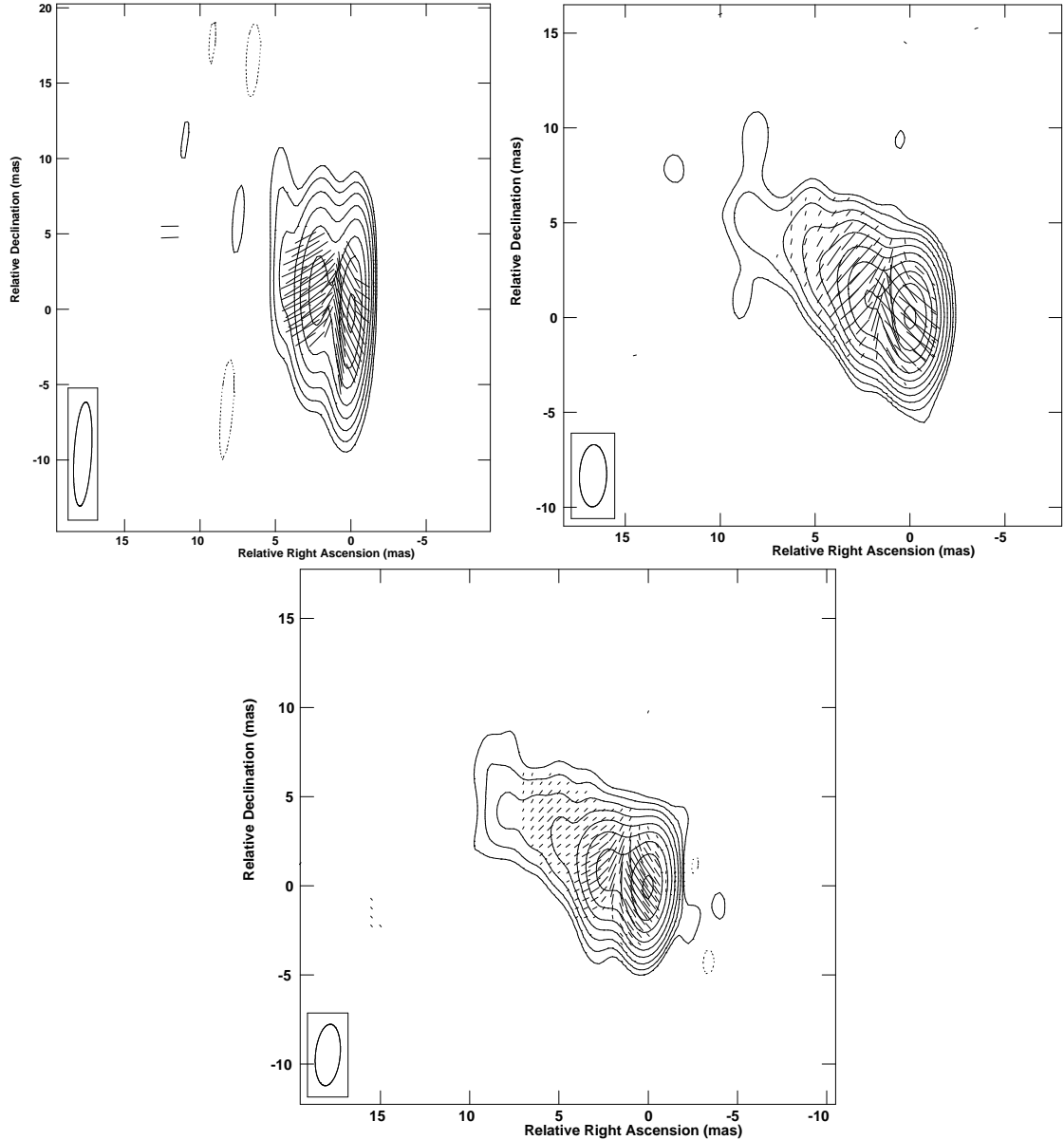}}
\caption{\small
Total-intensity images of the LBL 0829+046 with $\chi$ vectors superimposed. 
(Top left) Epoch 1993.15. 
Contours are --0.70, 0.70, 1.40, 2.80, 5.60, 11.20, 22.50, 45
and 90 per cent of the peak brightness of 469 mJy beam$^{-1}$,
$\chi$ vectors: 1 mas = 4 mJy beam$^{-1}$.
(Top right) Epoch 1995.53.
Contours are --0.2, 0.2, 0.35, 0.70, 1.40, 2.80, 5.60, 11.20, 22.50, 45
and 90 per cent of the peak brightness of 699.3 mJy beam$^{-1}$,
$\chi$ vectors: 1 mas = 5 mJy beam$^{-1}$.
(Bottom) Epoch 1998.49.
Contours are --0.35, 0.35, 0.70, 1.40, 2.80, 5.60, 11.20, 22.50, 45
and 90\% of the peak surface brightness of 441 mJy beam$^{-1}$,
$\chi$ vectors: 1 mas = 8 mJy beam$^{-1}$.}
\label{fig:0829}
\end{figure*}

\subsection{0829+046} 
0829+046 or OJ~049 is a $\gamma$-ray loud blazar \citep{Dondi95} which shows
rapid and large optical variability \citep{LillerLiller75}. 
The 1.4~GHz VLA images of \citet{AntonucciUlv85} and \citet{Giroletti04a}
show a two-sided structure with an extended and curved region of emission
to the southeast. Previous VLBI images show a VLBI jet extending to the
northeast \citep{Jorstad01}, clearly misaligned with the kpc-scale radio
structure.  Our VLBI images (Fig.~\ref{fig:0829}) reveal the rich polarization
structure of the VLBI jet, whose inferred $B$ field geometry has remained 
roughly constant over about five years. The predominant jet $B$ field is 
longitudinal to the jet. The polarization position angle for the knot K4 
changes dramatically over the
roughly five years covered by our observations, and seems to swing
to remain perpendicular to the VLBI jet as this component propagates
from the core (making the dominant $B$ field longitudinal essentially
throughout the jet). Both K3 and K4 show appreciable increases in the
degree of polarization accompanied by decreases in total intensity,
suggesting this is associated with the expansion of these components
as they evolve.

\begin{figure*}
\centerline{
\includegraphics[height=8cm]{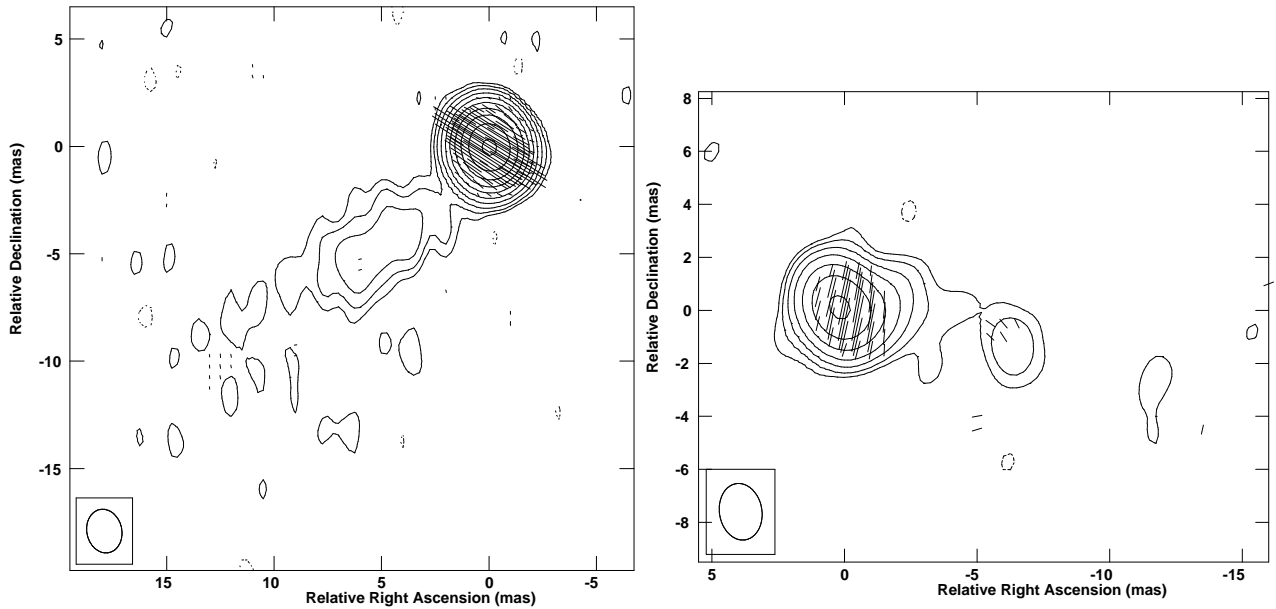}
\includegraphics[height=7cm]{bk55-1011-IP.ps}}
\caption{\small Total-intensity images with $\chi$ vectors superimposed.
(Left) The LBL 0925+504, epoch 1998.49.
Contours are --0.17, 0.17, 0.35, 0.70, 1.40, 2.80, 5.60, 11.20,
22.50, 45 and 90\% of the peak brightness of 433.3 mJy beam$^{-1}$,
$\chi$ vectors: 1 mas = 10 mJy beam$^{-1}$.
(Right) The HBL 1011+496, epoch 1998.49.
Contours are --1.4, 1.40, 2.80, 5.60, 11.20,
22.50, 45 and 90\% of the peak brightness of 93.8 mJy beam$^{-1}$,
$\chi$ vectors: 1 mas = 3 mJy beam$^{-1}$.}
\label{fig:0929}
\end{figure*}

\subsection{0925+504} 
The 1.4~GHz VLA image of \citet{Giroletti04a} shows a faint jet-like 
extension to the southeast, while the 8.4~GHz VLA map of \citet{Patnaik92}  
shows only the unresolved core. The source is quite core-dominated in our
VLBI image (Fig.~\ref{fig:0929}), but also shows a clear jet to the 
southeast, well aligned with
the structure observed by \citet{Giroletti04a}. The VLBI core displays the 
very high fractional polarization of $m_c=15\%$, while no polarization
was detected in the jet.

\subsection{1011+496} 
This object has been detected in TeV $\gamma$-rays by the MAGIC
telescope \citep{Albert07}. The 1.5~GHz VLA image of 
\citet{KollgaardPalma96} shows extended emission towards the west. 
Our VLBI image (Fig.~\ref{fig:0929}) shows a VLBI jet extending 
towards the west. Polarization was detected in the core, 
inner jet, and outer jet, $\sim$ 6 mas from the core. The $\chi$ 
vectors imply that the $B$ field is roughly aligned with 
the inner jet, but may become transverse in the outer jet. 

\begin{figure*}
\centering{
\includegraphics[height=7.5cm]{bk33-1101-IP.ps}
\includegraphics[height=7.5cm]{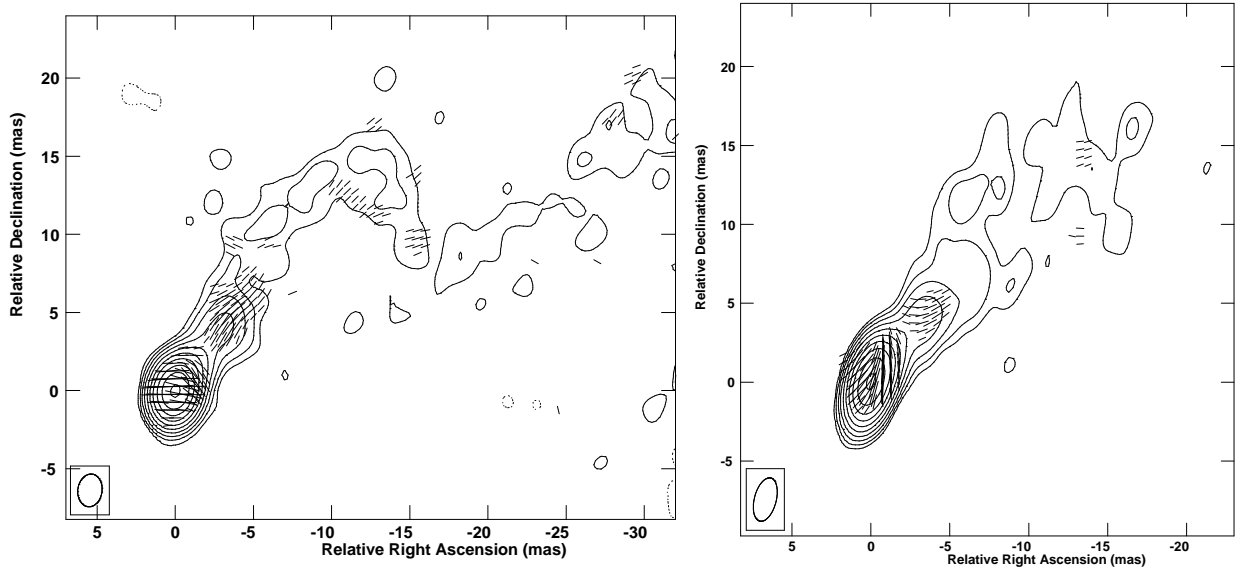}}
\caption{
Total-intensity images of the HBL 1101+384 (Mrk~421) with $\chi$ vectors 
superimposed. (Left) Epoch 1995.53.
Contours are --0.17, 0.17, 0.35, 0.70, 1.40, 2.80, 5.70, 11.50, 22.50, 45
and 90\% of the peak surface brightness of 326 mJy beam$^{-1}$,
$\chi$ vectors: 1 mas = 0.9 mJy beam$^{-1}$.
(Right) Epoch 1998.5.
Contours are  --0.17, 0.17, 0.35, 0.70, 1.40, 2.80, 5.70, 11.50, 22.50, 45
and 90\% of the peak surface brightness of 356.5 mJy beam$^{-1}$,
$\chi$ vectors: 1 mas = 1.8 mJy beam$^{-1}$.}
\label{fig:1101}
\end{figure*}

\subsection{1101+384} 
1104+382 or Mrk~421 is a well-known TeV $\gamma$-ray emitting blazar
\citep{MushotzkyBoldt78,SchwartzGriffiths79}. 
In the 0.5--10 keV band, this source is variable on time-scales ranging
from 14 hours to several days and occasionally exhibits large (by a 
factor of 10) X-ray outbursts characterized by a marked flattening of
the spectrum \citep{George88}.  The 1.4~GHz kpc-scale images of 
\citet{Kapahi79} and \citet{UlvestadJohnston83} show extended emission 
towards the northwest and northeast. The 22~GHz VLBP images of
\citet{PinerEdwards05} and 5--22~GHz VLBP images of \citet{Charlot06}
show the VLBI jet extending to the northwest, with the dominant inferred
$B$ field in the inner jet being transverse to the jet, although jet
regions with possibly oblique (i.e., neither aligned nor perpendicular 
to the jet) and longitudinal $B$ fields are also visible in the images
of \citet{Charlot06}.  Our new VLBI images (Fig.~\ref{fig:1101})
are qualitatively similar to
those of \citet{Charlot06}, with the predominant jet $B$ field being
transverse. Our images, especially the one from 1995, also show 
a region of faint, extended polarization $\sim$ 15--20~mas from the
core, with the $\chi$ vectors being roughly aligned with the direction back
towards the core; precisely the same structure is visible in the images
of \citet{Charlot06}, particularly the one for March 28, 1998. This
suggests that the observed total-intensity emission may represent 
only the edge of a much broader flow whose $B$ field is transverse
to the flow direction, as would be expected if we are seeing
the transverse component of a toroidal or helical $B$ field associated 
with the jet.
Overall, the variations in the polarizations
of individual components are appreciable, but not dramatic, being comparable
to the 5~GHz polarization variations observed on similar timescales for
individual VLBI components in LBLs \citep[e.g.,][]{Gabuzda94}.

\subsection{1133+704} 
\begin{figure*}
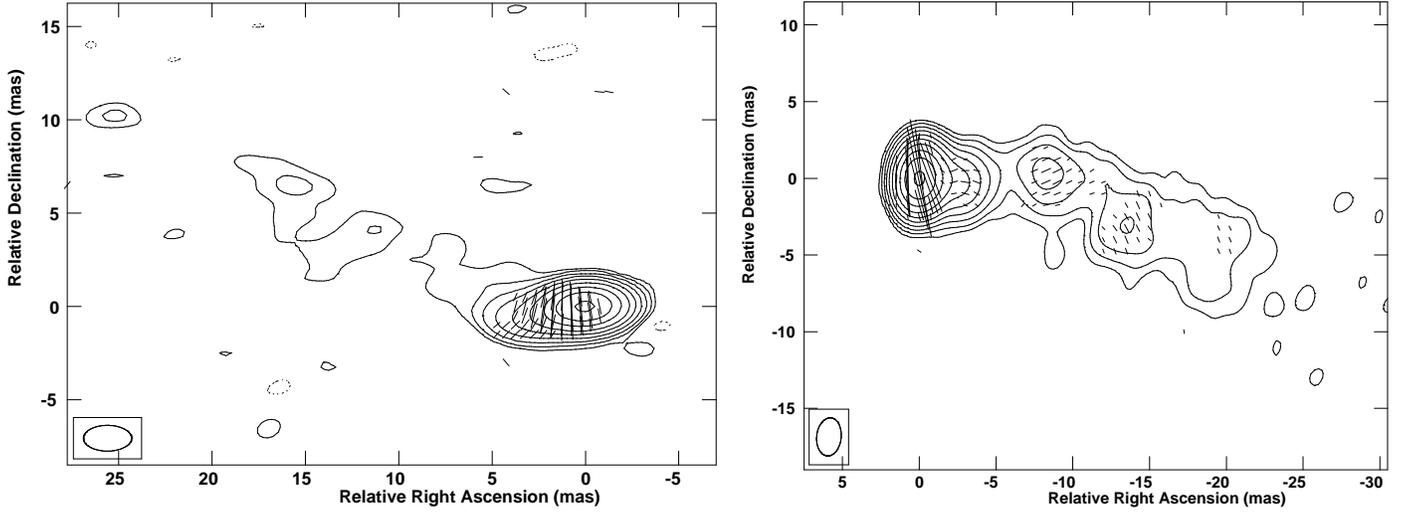

\centerline{
\includegraphics[height=7cm]{bk55-1133-IP.ps}
\includegraphics[height=7cm]{bk33-1147-IP.ps}}
\caption{\small Total-intensity images with $\chi$ vectors superimposed.
(Left) The HBL 1133+704, epoch 1998.49.
Contours are --0.35, 0.35, 0.70, 1.40, 2.80, 5.70, 11.50, 22.50, 45
and 90\% of the peak surface brightness of 98 mJy beam$^{-1}$,
$\chi$ vectors: 1 mas = 1.7 mJy beam$^{-1}$.
(Right) The LBL 1147+245, epoch 1995.53.
Contours are --0.17, 0.17, 0.35, 0.70, 1.40, 2.80, 5.6, 11, 23,
45 and 90\% of the peak surface brightness of 479.4 mJy beam$^{-1}$,
$\chi$ vectors: 1 mas = 2.5 mJy beam$^{-1}$.}
\label{fig:1133}
\end{figure*}

1133+704 is associated with the giant elliptical galaxy Mrk~180. 
X-ray emission was detected by the $Einstein$ and {\it HEAO-1} observatories 
\citep{Hutter80}, and the object is a candidate TeV BL~Lac \citep{Costamante02}.  
The kpc-scale radio structure has an asymmetric core-halo morphology 
\citep{AntonucciUlv85,LaurentMuehleisen93,Giroletti04a}. The previous global 
VLBI observations of \citet{Kollgaard96} show a jet to the southeast. A high degree 
of polarization was found in the inner jet, but with the corresponding $\chi$ vectors 
bearing no obvious relationship to the jet direction. Our new VLBI image 
(Fig.~\ref{fig:1133}) reveals polarized emission in both the core and jet, with the 
predominant implied $B$ field being longitudinal. We also detect a region
of polarization at the southern edge of the jet with $\chi \simeq 45\degr$,
whose origin is not clear.

\subsection{1147+245} 
This source has two remarkable properties that are quite unusual for BL~Lac 
objects: the radio structure at 1.4~GHz consists of a classical triple source 
\citep{AntonucciUlv85}, and the optical variability is rather small, with a total 
amplitude of only 0.77 mag \citep{Pica88}. The kpc-scale radio emission extends 
roughly in the north--south direction. The previous 5~GHz VLBP image of 
\citet{GabuzdaPushkarevCawthorne99} shows the VLBI jet extending to the west, 
misaligned with the large-scale structure. Only weak polarization was detected from 
the jet, corresponding to a transverse $B$ field. In our new VLBI image 
(Fig.~\ref{fig:1133}), polarization is detected throughout the jet,
with the $\chi$ vectors in the inner jet at and beyond $\sim$ 15 mas 
from the core being aligned with the jet, indicating transverse $B$ fields.  
The relative orientation of the $\chi$ vectors in another region of 
polarization $\sim 8$ mas from the core is not clear.

\begin{figure*}
\centering{
\includegraphics[height=8.3cm]{gk7-1219-I.ps}}
\includegraphics[height=8.1cm]{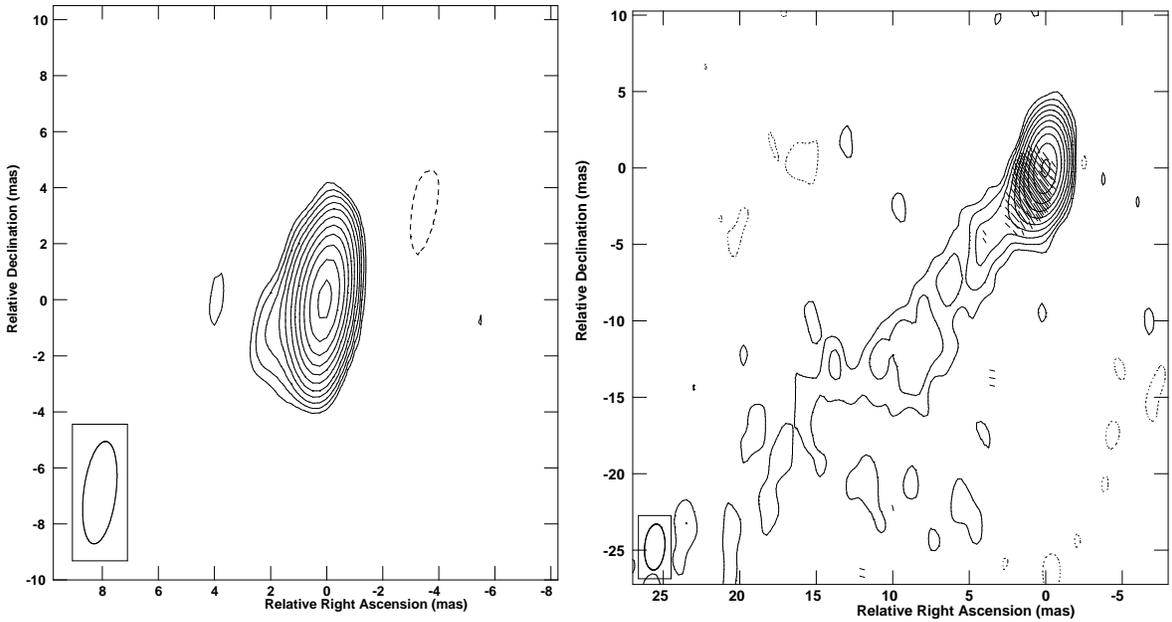}
\caption{\small
Total-intensity images of the HBL 1215+303 with $\chi$ vectors superimposed. 
(Left) Epoch 1993.15 (no polarization detected).  Contours are --2.8, 2.8, 
4.0, 5.6, 8, 11, 16, 23, 32, 45, 64 and 90\% of the peak surface brightness 
of 33.4 mJy beam$^{-1}$.  (Right) Epoch 1998.49.
Contours are  --0.17, 0.17, 0.35, 0.70, 1.40, 2.80, 5.70, 11.50, 22.50,
45 and 90\% of the peak surface brightness of 231 mJy beam$^{-1}$,
$\chi$ vectors: 1 mas = 2 mJy beam$^{-1}$.}
\label{fig:1215}
\end{figure*}

\subsection{1215+303} 
The 1.5~GHz VLA image of \citet{LaurentMuehleisen93} shows an unresolved
core with a hint of extended emission towards the southwest. The core-to-lobe
ratio is $\geq 14$, the largest in the {\it HEAO-1} sample. Our 1993 image shows 
a short jet in PA$\sim 135\degr$, and our 1998 image reveals a well-defined 
straight jet in this same position angle (Fig.~\ref{fig:1215}). Polarization was 
detected only at the latter epoch, in both the core and inner jet. The inferred 
jet $B$ field is aligned with the jet. The jet component K5 was roughly 22\% polarized
at epoch 1998.49, when the total intensity of this feature was about a factor
of eight higher than at our earlier epoch, 1993.15. If K5 were comparably
polarized in 1993.15, it would have had a polarized flux of only about
0.7--0.8~mJy, making the absence of a polarization
detection at the earlier epoch unsurprising.

\subsection{1227+255} 
\begin{figure*}
\centerline{
\includegraphics[height=8.0cm]{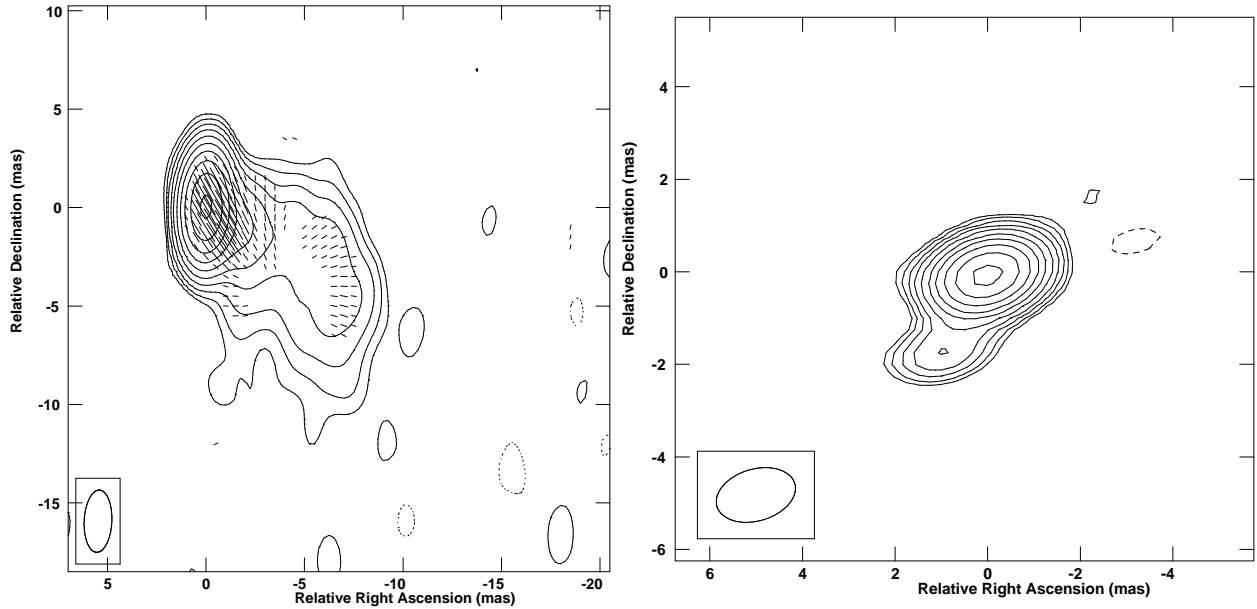}
\includegraphics[height=8.0cm]{gk7-1235-I.ps}}
\caption{\small Total-intensity images with $\chi$ vectors superimposed. 
(Left) The HBL 1227+255, epoch 1998.49.
Contours are --0.17, 0.17, 0.35, 0.70, 1.40, 2.80, 5.60, 11.20, 22.50, 45
and 90\% of the peak surface brightness of 183.7 mJy beam$^{-1}$,
$\chi$ vectors: 1 mas = 2.5 mJy beam$^{-1}$.
(Right) The HBL 1235+632, epoch 1993.15 (no polarization detected).
Contours are --4, 4, 5.6, 8, 11, 16, 23, 32, 45, 64, 90\% of the
peak surface brightness of 13.8 mJy beam$^{-1}$.}
\label{fig:1227}
\end{figure*}

This BL~Lac object was identified during the correlation of the ROSAT All-Sky
Survey (RASS) with the Hamburg Quasar Survey \citep{NassBade96}. 
Our VLBI image (Fig.~\ref{fig:1227}) shows a jet that initially emerges to 
the southwest, then curves toward the south. Polarization was detected in both the 
pc-scale core and jet. The $\chi$ vectors in the inner jet are roughly aligned with 
the jet direction, implying a transverse $B$ field, and a `sheath' of longitudinal 
$B$ field is observed along the northern edge of the jet, where the jet appears 
to be bending.

\subsection{1235+632} 
The 1.5~GHz VLA image of \citet{LaurentMuehleisen93} does not reveal
any extended structure. Our VLBI image (Fig.~\ref{fig:1227}) shows a 
pc-scale jet extending in PA$\sim 140\degr$. No polarization was detected.

\begin{figure*}
\centerline{
\includegraphics[height=8.0cm]{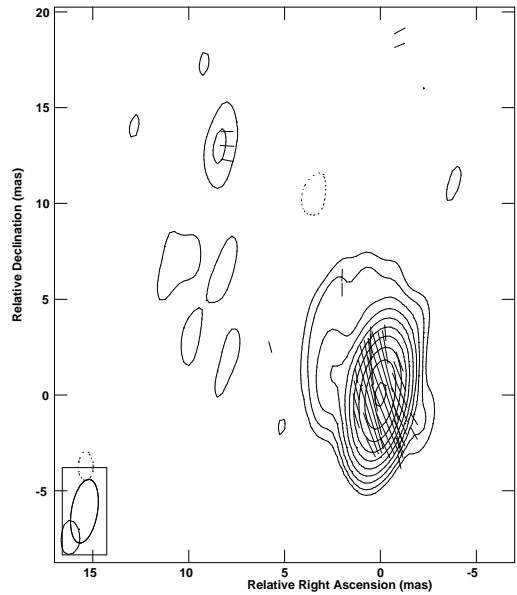}}
\caption{\small Total-intensity image of the LBL 1553+113, epoch 1998.49,
with $\chi$ vectors superimposed.
Contours are --0.19, 0.19, 0.35, 0.70, 1.40, 2.80, 5.60, 11, 23, 45
and 90\% of the peak surface brightness of 189.7 mJy beam$^{-1}$,
$\chi$ vectors: 1 mas = 1.5 mJy beam$^{-1}$.}
\label{fig:1553}
\end{figure*}

\subsection{1553+113}
The 1.4~GHz VLA image of \citet{Rector03} shows a faint lobe south 
of the core, with a weak ``hot spot'' in PA=160$\degr$. The 5~GHz VLBA 
map however shows a jet extending to the northeast in PA$=48\degr$, giving a 
large misalignment of $\Delta$PA=112$\degr$ \citep{Rector03}. Our VLBI map 
(Fig.~\ref{fig:1553}) shows a dominant core and a short jet in PA$\sim 45\degr$. 
There may be fainter jet emission further from the core in this same direction, 
and also possibly a faint extension on the counterjet side. Polarization was detected
in both the core and inner jet, but the orientation of the $\chi$ vectors relative 
to the jet is unclear.

\begin{figure*}
\centering{
\includegraphics[height=7.6cm]{bk33-1727-IP.ps}
\includegraphics[height=7.6cm]{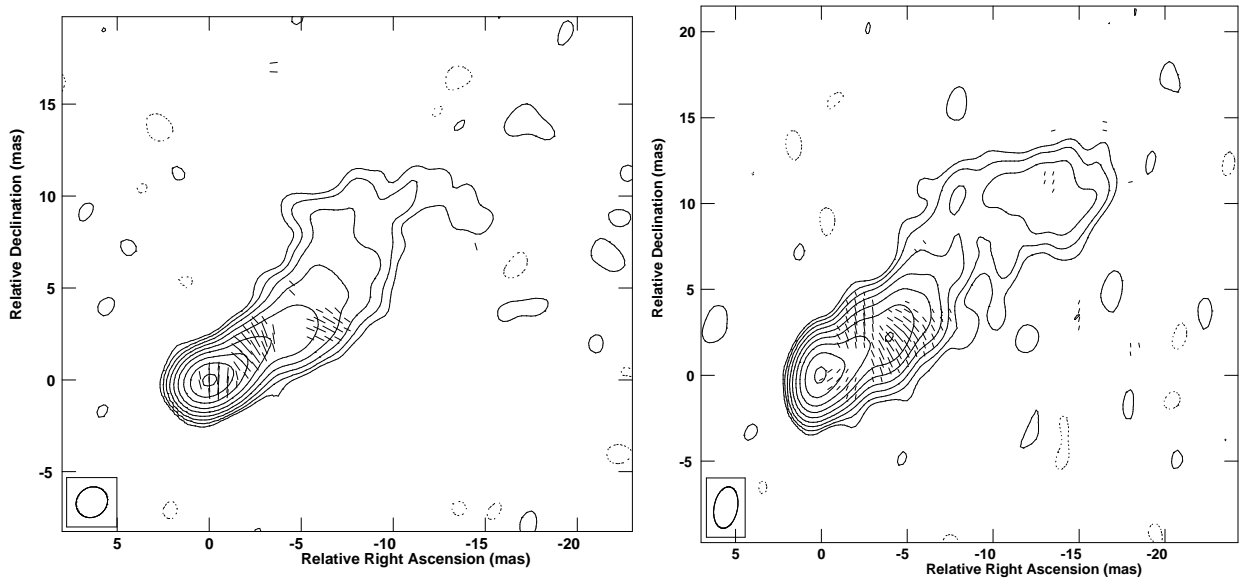}}
\caption{\small
Total intensity image of the HBL 1727+502 with $\chi$ vectors superimposed.
(Left) Epoch 1995.53.
Contours are --0.35, 0.35, 0.70, 1.40, 2.80, 5.60, 11, 23, 45
and 90\% of the peak surface brightness of 77.2 mJy beam$^{-1}$,
$\chi$ vectors: 1 mas = 1 mJy beam$^{-1}$.
(Right) Epoch 1998.49.
Contours are --0.35, 0.35, 0.70, 1.40, 2.80, 5.60, 11.20,
22.50, 45 and 90\% of the peak brightness of 64.2 mJy beam$^{-1}$,
$\chi$ vectors: 1 mas = 1.8 mJy beam$^{-1}$.}
\label{fig:1727}
\end{figure*}

\subsection{1727+502} 
1727+502 is a candidate TeV BL~Lac \citep{Costamante02}.  
The kpc-scale radio structure consists of an unresolved core and diffuse, 
asymmetrical halo, with evidence for a jet emerging to the northwest 
\citep{WardleMoore84,Giroletti04a}. The previous 5~GHz VLBP image of
\citet{Kollgaard96} show a jet extending to the northwest, roughly aligned
with the kpc-scale structure. Polarization was detected in several 
jet components, with the implied $B$ field being either transverse or
lontudinal in the inner jet, depending on interpretation, then becoming
clearly longitudinal further from the core. Our new VLBI images 
(Fig.~\ref{fig:1727}) clearly show regions of polarization corresponding to 
longitudinal $B$ fields offset from the jet ridge line, forming a sort of 
`sheath' of polarization. The 1998 image also shows a region of transverse $B$ 
field in the inner jet, thus overall a `spine-sheath'-like polarization structure.
There were no dramatic changes in the polarization properties of individual 
VLBI components in the roughly three years between our two epochs.

\subsection{1741+196} 
\begin{figure*}
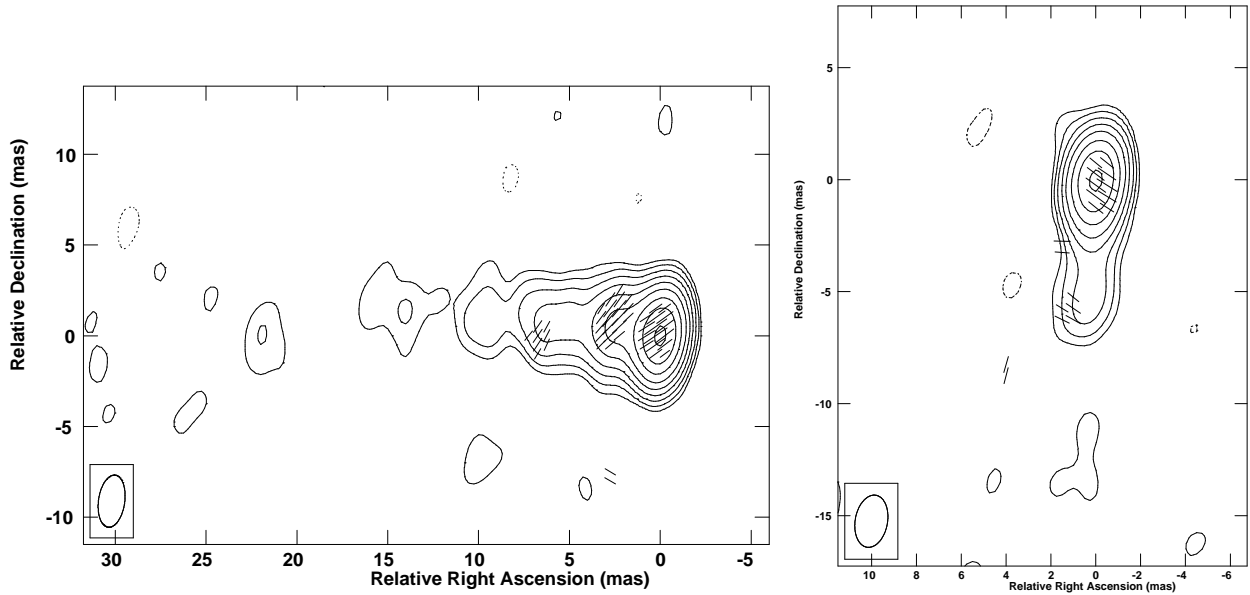

\centerline{
\includegraphics[height=7.0cm]{bk55-1743-IP.ps}
\includegraphics[height=8.0cm]{bk55-1745-IP.ps}}
\caption{\small 
Total-intensity images with $\chi$ vectors superimposed.
(Left) The HBL 1741+196, epoch 1998.49, 
Contours are --0.35, 0.35, 0.70, 1.40, 2.80, 5.60, 11.20, 22.50, 45
and 90\% of the peak surface brightness of 84.3 mJy beam$^{-1}$,
$\chi$ vectors: 1 mas = 1.3 mJy beam$^{-1}$.
(Right) The HBL 1743+398, epoch 1998.49.
Contours are --0.70, 0.70, 1.40, 2.80, 5.60, 11, 23, 45
and 90\% of the peak surface brightness of 49.3 mJy beam$^{-1}$,
$\chi$ vectors: 1 mas = 0.9 mJy beam$^{-1}$.}
\label{fig:1743}
\end{figure*}

VLA snapshot observations of this source \citep[see][]{Perlman96} show a jet in 
PA$\sim 90\degr$. The 5~GHz VLBA image of \citet{Rector03} shows a collimated
straight jet extending to the east in PA$=86\degr$, well aligned with the
VLA jet.  Our VLBI map (Fig.~\ref{fig:1743}) clearly shows a well-defined 
straight jet extending in PA$\sim 90\degr$. Polarization was detected in the core and 
two jet components, but the $\chi$ vectors bear no obvious relationship
to the direction of the jet structure.

\subsection{1743+398} 
This BL~Lac object lies close to the center of a moderately massive galaxy 
cluster \citep{Nilsson99}. The 1.4~GHz VLA image of this source shows a highly 
distorted FRI morphology; the jets lie in the northwest and southeast 
directions, and show sharp bends \citep{RectorGabuzda03}. Our VLBI 
image (Fig.~\ref{fig:1743}) shows a jet pointing nearly to the south. 
There is evidence for a `sheath' of longitudinal $B$ field at the 
eastern edge of the jet.

\begin{figure*}
\centerline{
\includegraphics[height=8.0cm]{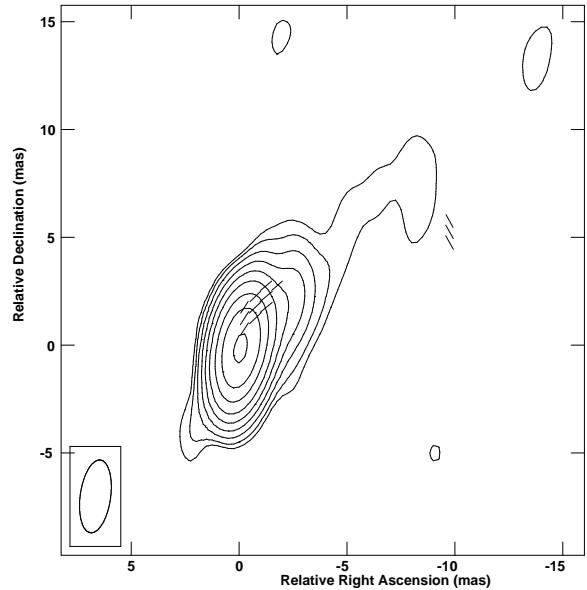}}
\caption{\small
Total-intensity images of the Sy-1 2201+044, epoch 1998.49, with $\chi$ vectors 
superimposed. Contours are --0.4, 0.4, 0.70, 1.40, 2.80, 5.60, 11, 23, 45
and 90\% of the peak surface brightness of 162.3 mJy beam$^{-1}$,
$\chi$ vectors: 1 mas = 0.9 mJy beam$^{-1}$.}
\label{fig:2201}
\end{figure*}

\subsection{2201+044} 
Many radio observations have mapped the intricate extended structure in
this Seyfert-1 galaxy at different frequencies and resolutions 
\citep{UlvestadJohnston84,VanGorkom89}. The 1.5~GHz observations of 
\citet{LaurentMuehleisen93} revealed a broad jet extending towards the northwest 
in PA$\sim -50\degr$ and faint extended emission to the east. The previous VLBI 
image of \citet{Kollgaard96} showed a compact jet structure in PA$\sim -42\degr$, 
aligned with the dominant kpc-sale structure; no pc-scale polarization was 
detected. Our new VLBI map (Fig.~\ref{fig:2201}) shows a jet in the same 
position angle, $i.e., \sim -40\degr$. Polarization was detected only in 
the jet, with the inferred $B$ field being perpendicular to the jet direction.

\section{VLBI properties of LBLs and HBLs at 6 cm}

\begin{figure*}
\centerline{
\includegraphics[height=7.0cm]{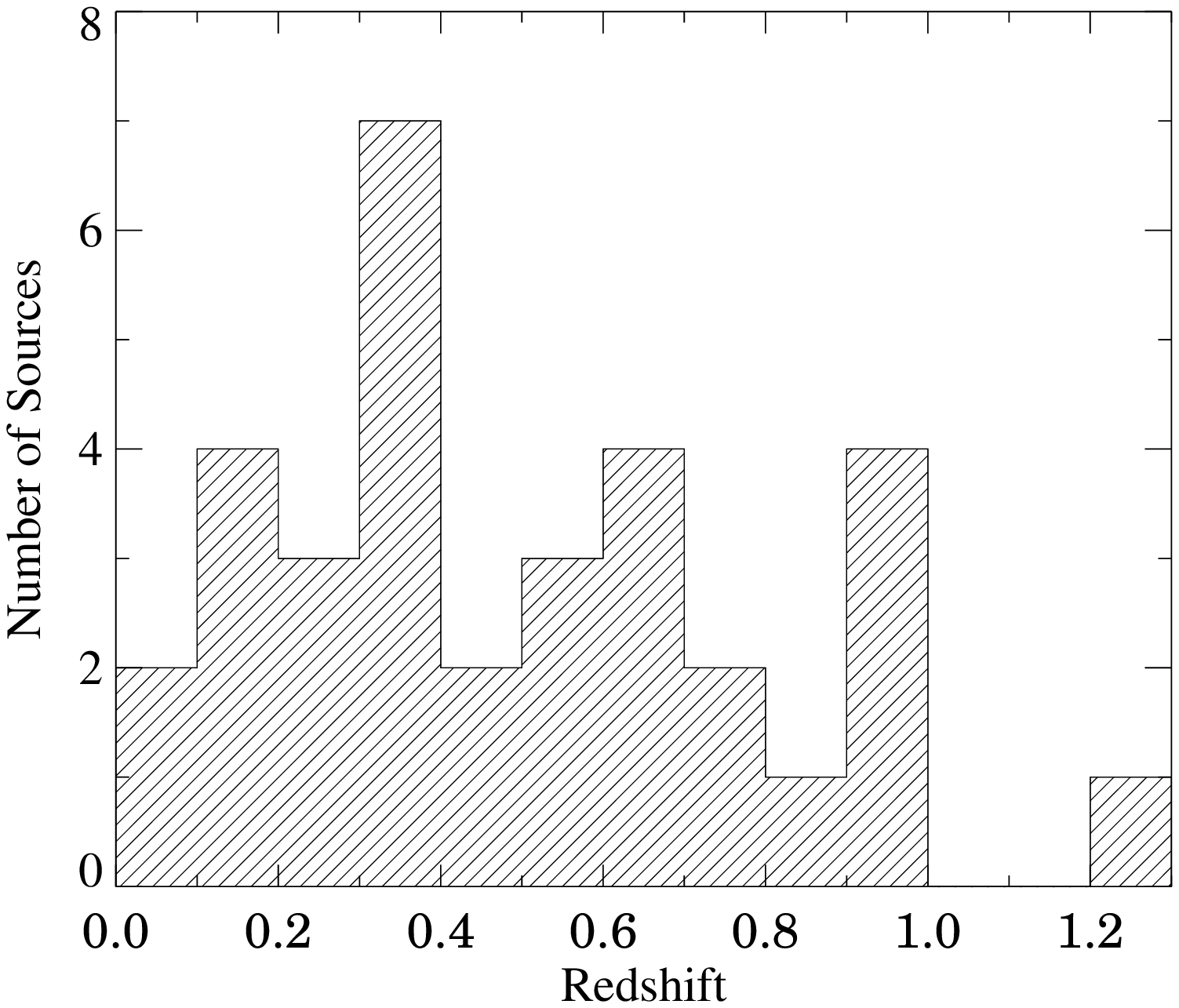}
\includegraphics[height=7.0cm]{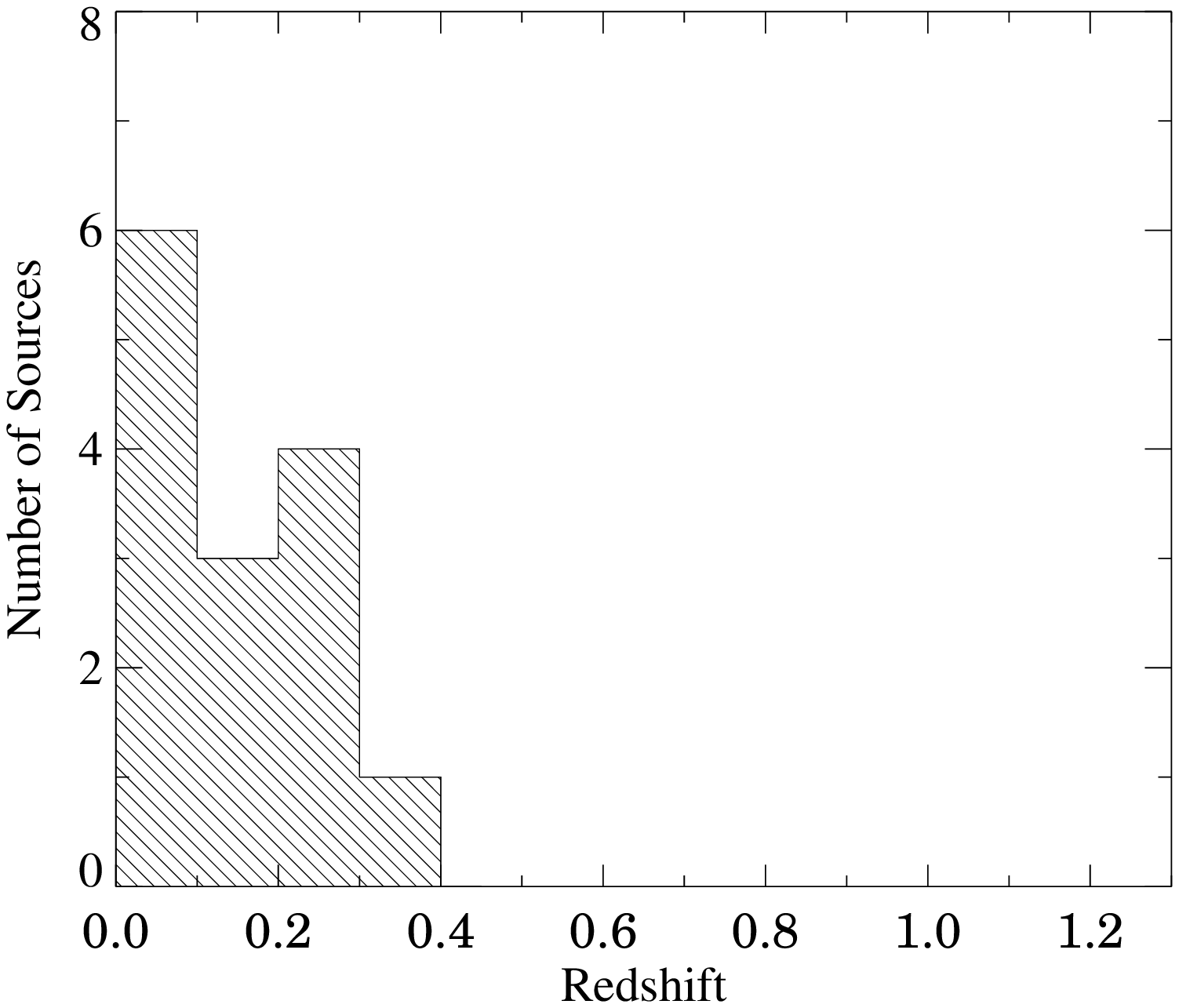}}
\caption{The redshift distributions of LBLs (left) and HBLs (right). 
The LBLs are primarily 1-Jy BL~Lacs \citet{Gabuzda00} but 
include four LBLs from our {\it HEAO-1}+RGB sample while the HBLs belong to 
our {\it HEAO-1}+RGB sample and include Mrk~501 from the 1-Jy sample.}
\label{fig:redshift}
\end{figure*}

The data for the LBLs, which belong primarily to the 1-Jy sample
\citep{KuehrSchmidt90}, have been obtained from \citet{Gabuzda00}. Data
for four additional LBLs belonging to our {\it HEAO-1}+RGB sample have
been included in the analysis. Of these, 1147+245 also belongs to the 
1-Jy sample. The data for the HBLs comes primarily
from our {\it HEAO-1}+RGB sample. The redshift distribution for the LBL
and HBL samples is presented in Fig.~\ref{fig:redshift}.
The distributions of the core fractional polarization, the inner-jet 
($r < 10$~pc from the core) 
fractional polarization, the difference between the core polarization 
angle and the jet direction $|\chi_c-\theta|$ and the difference between the
inner jet polarization angle and the jet 
direction $|\chi_j-\theta|$ for LBLs and HBLs are presented in 
Figs.~\ref{fig:m_c}, \ref{fig:chi_c}, \ref{fig:m_j}, and \ref{fig:chi_j}, 
respectively. Table~\ref{tab:versus} tabulates the values for these parameters. 
Note that multiple values for a single source are included in the histograms if 
they fall in different bins, but not if they fall in the same bin.
The Sy-1 galaxy 2201+044 was excluded from the analysis.

\begin{figure*}
\centerline{
\includegraphics[height=7.0cm]{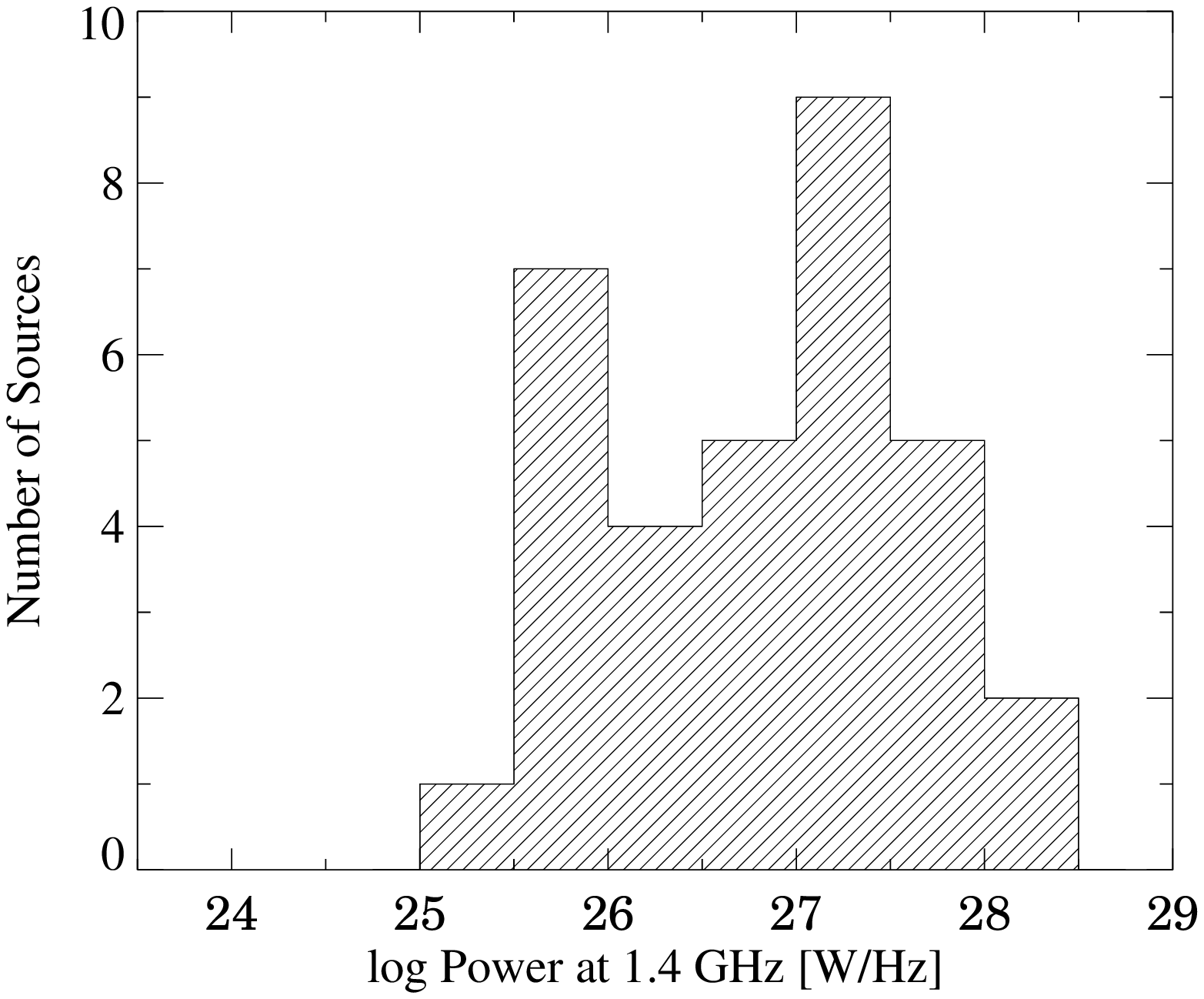}
\includegraphics[height=7.0cm]{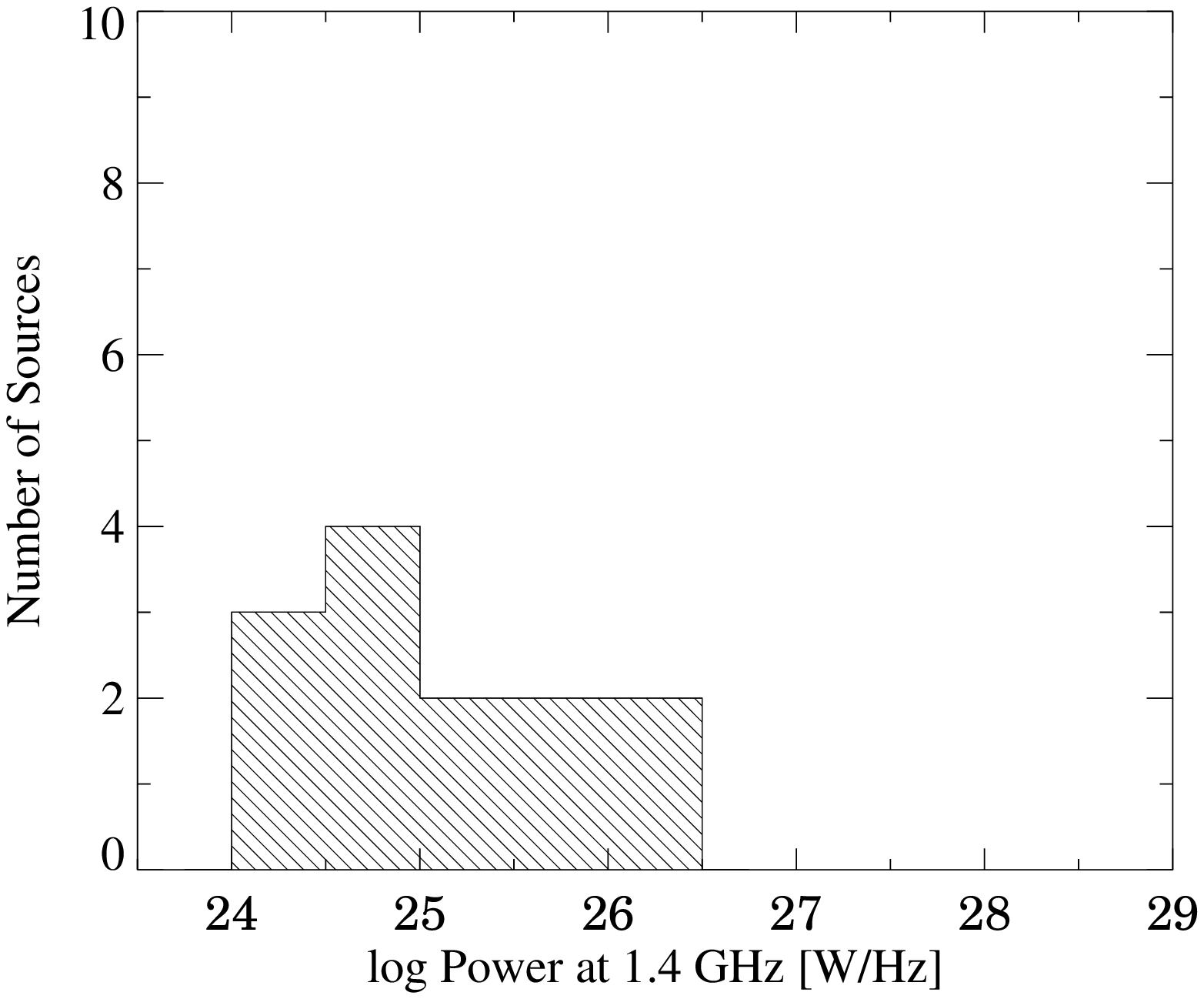}}
\caption{The total radio power at 1.4~GHz for LBLs (left) and HBLs (right). 
The LBLs are primarily 1-Jy BL~Lacs \citet{Gabuzda00} but 
include three LBLs from our {\it HEAO-1}+RGB sample while the HBLs belong to 
our {\it HEAO-1}+RGB sample and include Mrk~501 from the 1-Jy sample.
The 1.4~GHz radio data was obtained from the NASA/IPAC Extragalactic Database. }
\label{fig:power}
\end{figure*}

\begin{table}
\begin{center}
\caption{HBL Core and Jet polarisation properties}
\begin{tabular}{cllclccccccc}\hline\hline
Source & $m_c$ & $m_j$ [Jet Component] & $|\chi_c-\theta|$ &$|\chi_j-\theta|$
[Jet Component]\\
       & (\%)  & (\%)  & (deg)             & (deg)           \\\hline
0414+009 & ... & ...                 & ... & ...   \\	 
0652+426 & ... & 3.8                 & ... & 42  \\	 
0706+592 & ... & ...                 & ... & ... \\	 
0749+540 & 9.7 & ...                 & ... & ... \\	 
0829+046 & 3.2 & 4.0 [K4], 5.8 [K3]  & 30  & 50 [K4], 80 [K3] \\	 
         & 2.9 & 6.1 [K5], 5.9 [K4]  & 3   & 51 [K5], 88 [K4] \\	 
         & 4.4 & 10.9 [K6], 4.9 [K5] & 21  & 62 [K6], 65 [K5] \\	 
0925+504 & 14.2& ...                 & 69  & ...   \\	 
1011+496 & 4.7 & 3.5                 & 82  & 80  \\	 
1101+384 & 1.5 & 7.7 [K8], 27.2 [K6] & 55  & 55 [K8], 6 [K6] \\	 
         & 1.2 & 15.4 [K5], 9.8 [K3] & 14  & 37 [K5], 30 [K3]\\	 
1133+704 & ... & 28.6 [K6], 14.5 [K5]& ... & 81 [K6], 34 [K5]  \\	 
1147+245 & 4.3 & 5.3                 & 73  & 12  \\ 	 
1215+303 & ... & 22.0                & ... & 64  \\	 
1227+255 & 2.5 & 22.0 [K4], 10.6 [K3]& 25  & 22 [K4], 60 [K3]  \\	 
1553+113 & 2.3 & 14.7                & 30  & 50  \\	 
1727+502 & 1.6 & 5.3 [K3], 8.8 [K2], 3.7 [K1]  & 43  & 90 [K3], 61 [K2], 69 [K1]  \\	 
         & ... & 2.0 [K5], 10.6 [K4], 11.6 [K3]& ... & 20 [K5], 61 [K4], 75 [K3]  \\	 
1741+196 & 1.1 & 9.7 [K5], 10.9 [K4]           & 52  & 67 [K5], 66 [K4]  \\	 
1743+398 & 1.5 & ...                           & 54  & ...   \\	 
2201+044 & ... & 2.5                           & ... & 8  \\\hline
\label{tab:versus}
\end{tabular}
\end{center}
{Notes $-$ 0829+046: there is an extended region with inferred
longitudinal $B$-field further from the core. 
1101+384: clear evidence for extended regions with $B$-field
perpendicular to the jet direction slightly further from the core.
1147+245: the inferred $B$-field appears to remain perpendicular
to the local jet direction further from the core as the jet bends.
1227+255: evidence for a sheath of longitudinal $B$-field 
further from the core. 1727+502: evidence for a sheath of longitudinal 
$B$-field at edges of the jet.}
\end{table}

Table~\ref{tab:KS} summarises the results of the comparison
between LBLs and HBLs. It has the following columns:
Col.~(1) the property being compared, (2) the corresponding two-sided 
Kolmogorov--Smirnov (K--S) statistic, which specifies here the maximum deviation 
between the cumulative distributions for the LBLs and HBLs, (3) the 
significance level of the K--S statistic, i.e., the probability that
the cumulative distributions for the LBLs and HBLs are intrinsically 
similar, and (4) ``YES" and ``NO" indicate if the differences between 
the properties of the two BL~Lac populations are statistically 
significant or not, while YES? indicates a difference that is significant at 
the 2$\sigma$ level.

\subsection{Parsec-scale core polarization}

\begin{figure*}
\centerline{
\includegraphics[height=7.0cm]{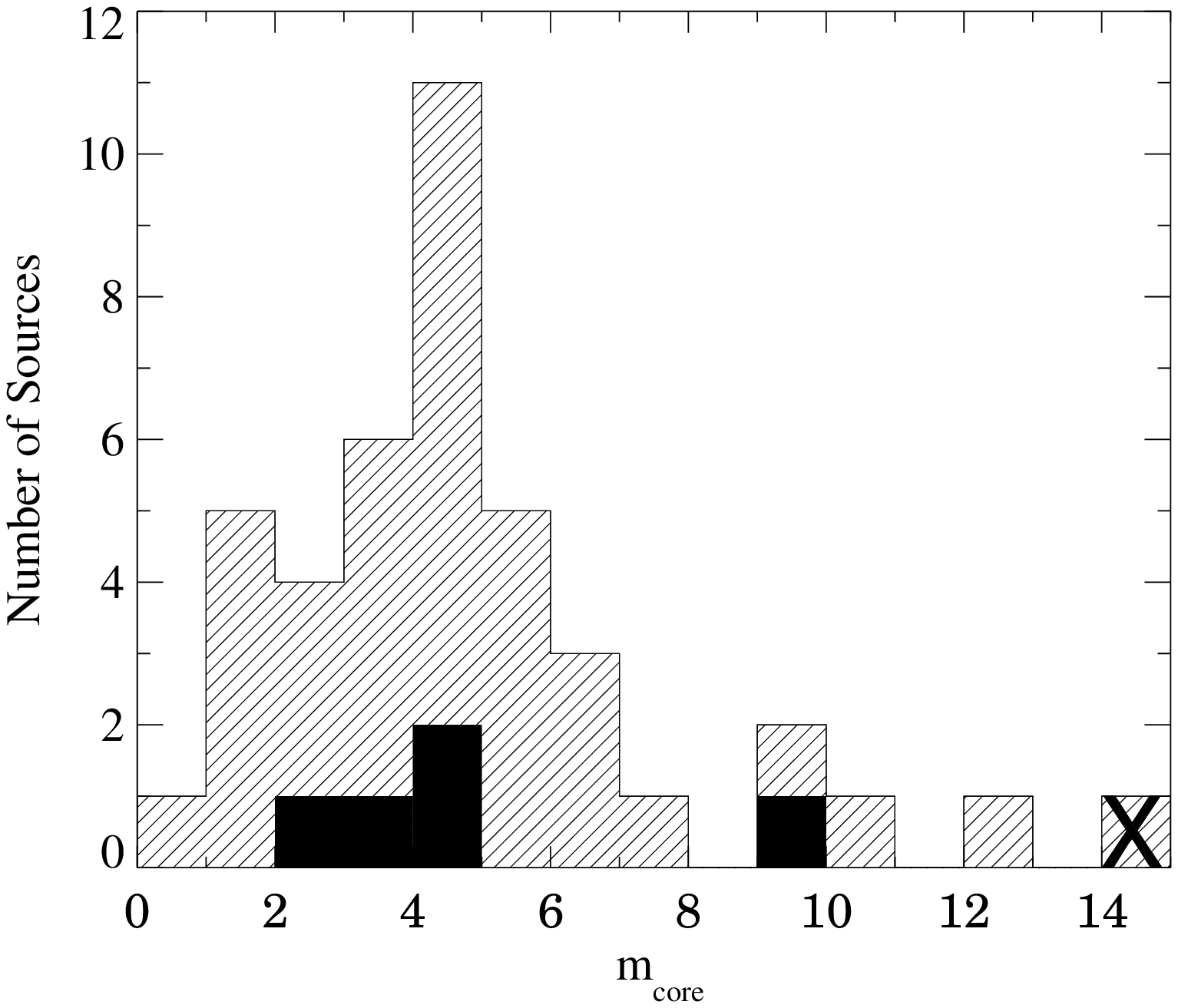}
\includegraphics[height=7.0cm]{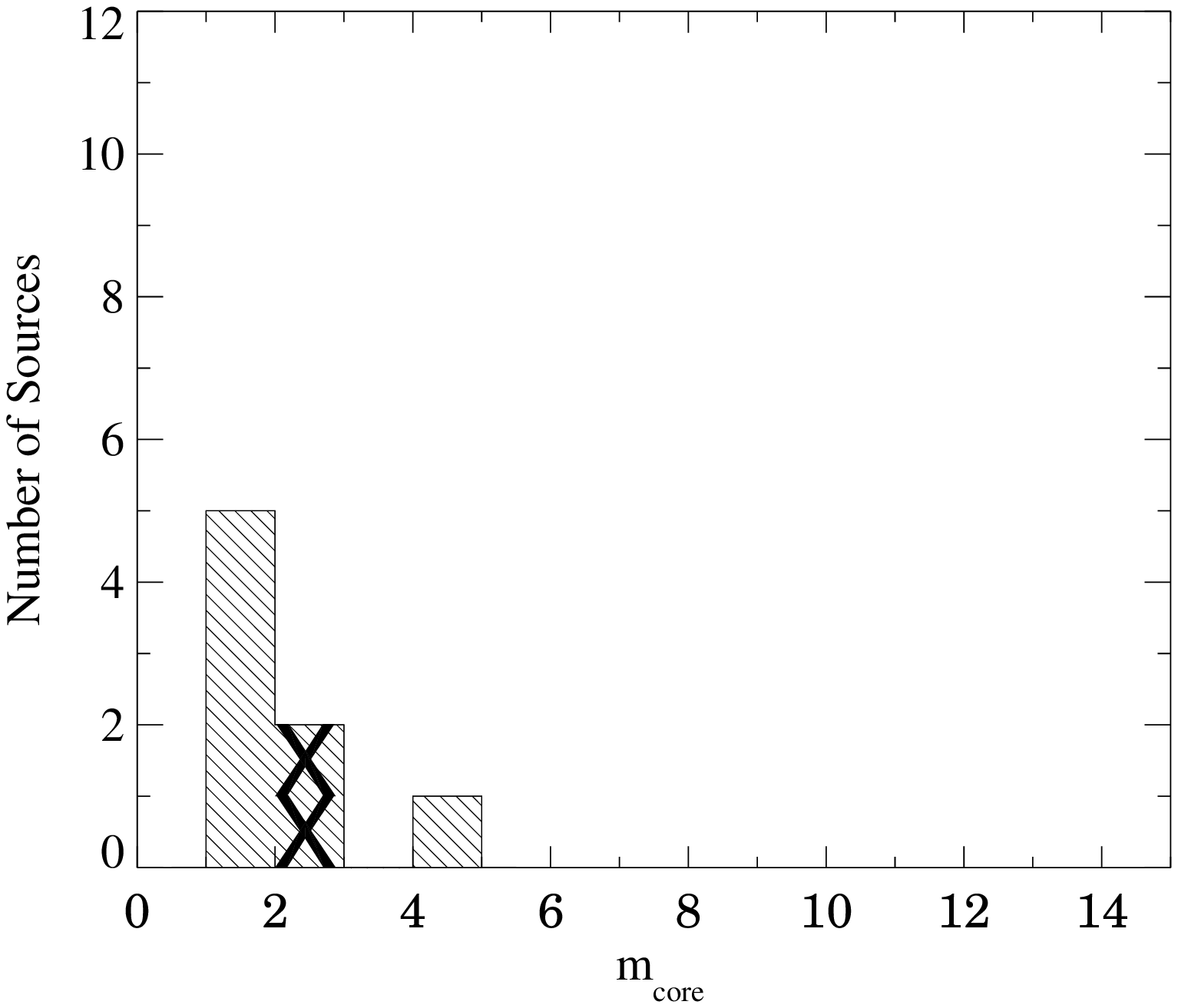}}
\caption{Distributions of core fractional polarization for the 1-Jy
LBLs (left) and HBLs (right). The shaded regions mark the LBLs belonging
to our {\em HEAO-1}+RGB sample. X denotes a BL~Lac that is alternately
classified as an IBL by \citet{Nieppola06}.}
\label{fig:m_c}
\end{figure*}

We have detected pc-scale polarized emission in all but three 
of the high-energy-peaked BL~Lacs in our {\it HEAO-1}+RGB sample.
We find that the degree of polarization in the core, $m_c$, is typically
less than 3\% in the HBLs, with only one HBL (1011+496) having
$m_c\ge4\%$ (see Fig.~\ref{fig:m_c}). 
On the other hand, LBLs typically have $m_c>3\%$ with the
core polarizations occasionally exceeding 10\% (Fig.~\ref{fig:m_c}). 
It is interesting that two of the X-ray selected BL Lac objects for which
results are presented here have among the highest core fractional polarizations
ever measured $-$ 0749+540 ($m_c\sim9\%$) and 0925+504 ($m_c\sim15\%$); however,
these prove to be LBLs, so that their core fractional polarizations fall
within the previously observed range for this class. 

A two-sided Kolmogorov--Smirnov test on the distribution of $m_c$ for the
HBL and LBL subclasses indicates only a 0.3\% probability that the
two datasets come from the same parent population (Table~\ref{tab:KS}).
The tendency for the LBLs to have higher pc-scale core fractional polarization 
than the HBLs has also been observed in the kpc-scale cores of these 
BL~Lac objects \citep{Kollgaard96}.
Although both LBLs and HBLs have relatively modest 
redshifts compared to the quasars, there is a significant difference
in the redshift distribution of LBLs and HBLs (Fig.~\ref{fig:redshift}). 
The K--S test indicates that the probability that the two datasets come from
the same parent population is only around 0.02\% (Table~\ref{tab:KS}).
This would result in the VLBI observations probing somewhat different spatial 
scales in the two BL~Lac subclasses. Therefore there is a  
possibility that, on average, the radio cores of LBLs include greater 
contributions from the inner radio jets than do the radio cores of HBLs, 
resulting in higher average core fractional polarizations. 
However, when we tested this we did not find any correlation between core 
fractional polarization and redshift for the LBLs 
(Spearman rank test; $p$=0.9). Further, this idea
could not simultaneously explain the origin of the lower core fractional 
polarizations in quasars compared to LBLs \citep{Gabuzda00}, 
since quasars tend to typically have {\em higher} redshifts than BL~Lac objects. 
Based on significant differences in the total and extended power,
as discussed later in \S~7, we infer the low $m_c$ in HBLs to be
due to their intrinsically weaker radio cores.

\begin{figure*}
\centerline{
\includegraphics[height=7.0cm]{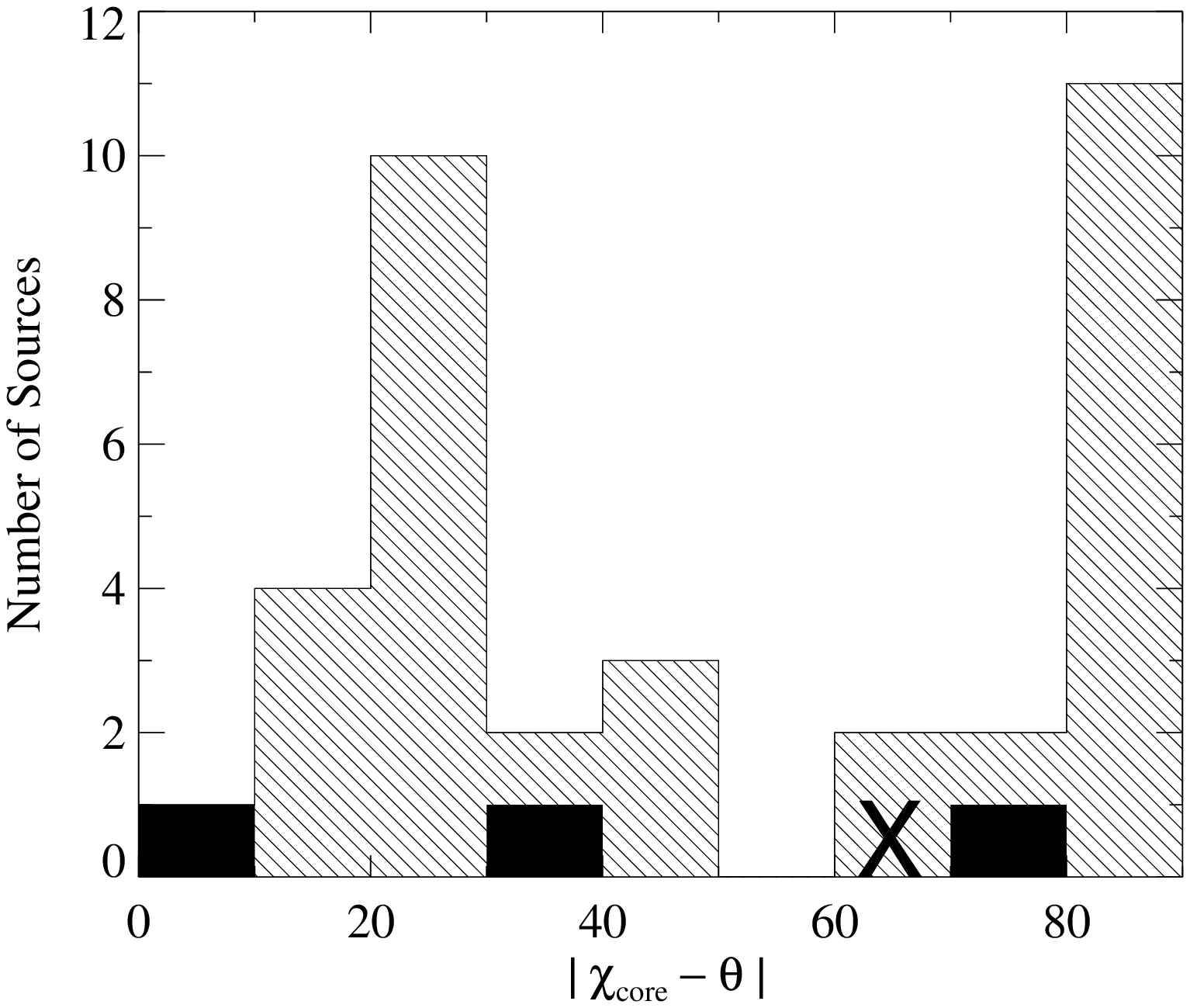}
\includegraphics[height=7.0cm]{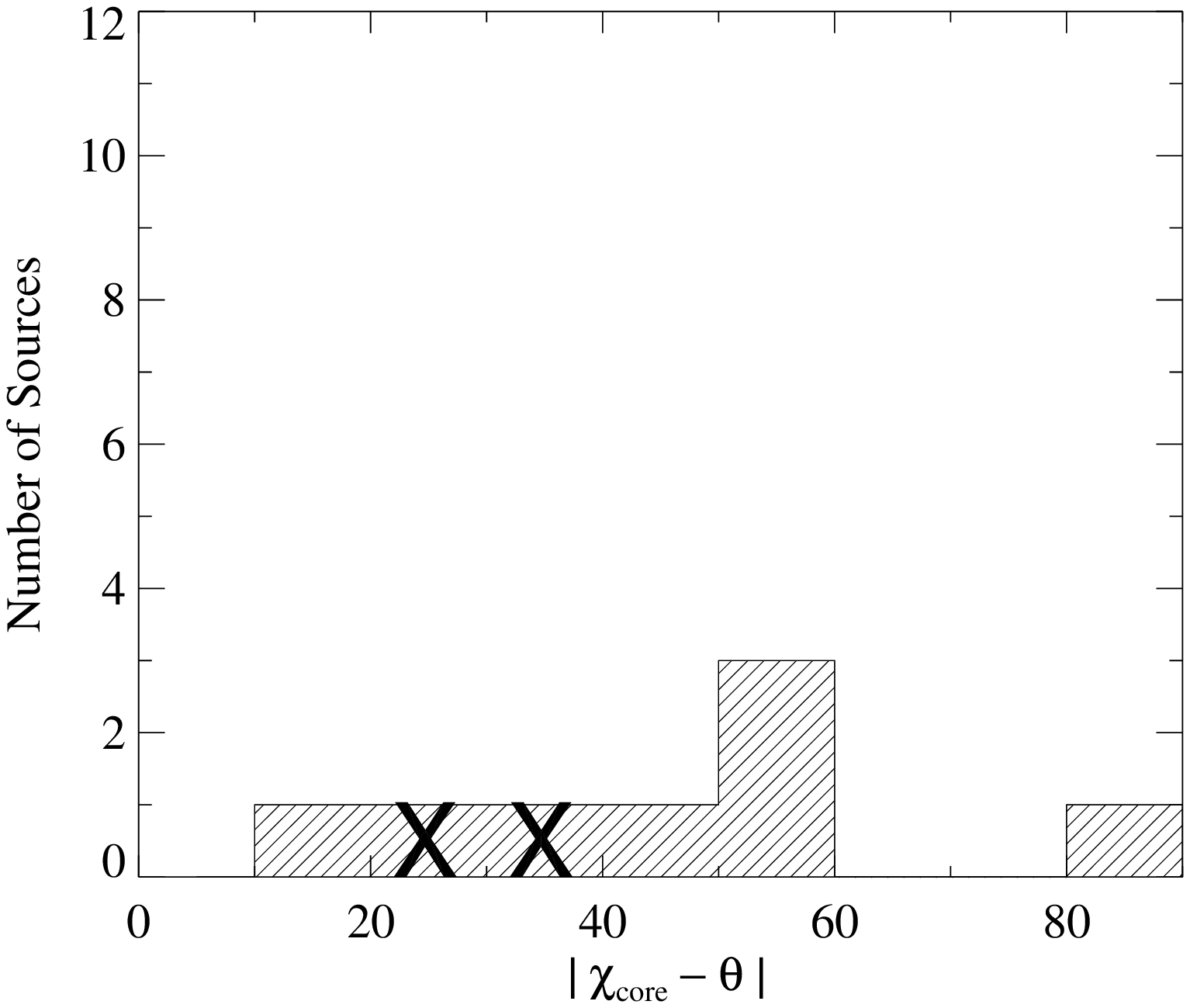}}
\caption{Distributions of offset between the core polarization angle
$\chi_c$  and the direction of the VLBI jet for the 1-Jy LBLs (left) and HBLs
(right). The distribution is bimodal for LBLs
with preferred values close to 0$^\circ$ and 90$^\circ$, while the
HBLs seem to show the whole range of angles. The shaded regions mark the 
LBLs belonging to our {\em HEAO-1}+RGB sample. X denotes a BL~Lac that 
is alternately classified as an IBL by \citet{Nieppola06}.}
\label{fig:chi_c}
\end{figure*}

We find that the polarization-angle orientations in the pc-scale cores 
relative to the inner VLBI jet direction are not statistically 
significantly different 
in HBLs and LBLs (see Fig.~\ref{fig:chi_c} and Table~\ref{tab:KS}). 
However, while in LBLs there is a clear tendency for $\chi_c$ to lie either 
parallel or perpendicular to the jet direction \citep[see][]{Gabuzda00},
such a behaviour is not observed in HBLs.
Optical depth effects as discussed by \citet{Gabuzda03} could however
be contributing to the differences in the observed EVPA distributions.

\subsection{Parsec-scale jet polarization}

\begin{figure*}
\centerline{
\includegraphics[height=7.0cm]{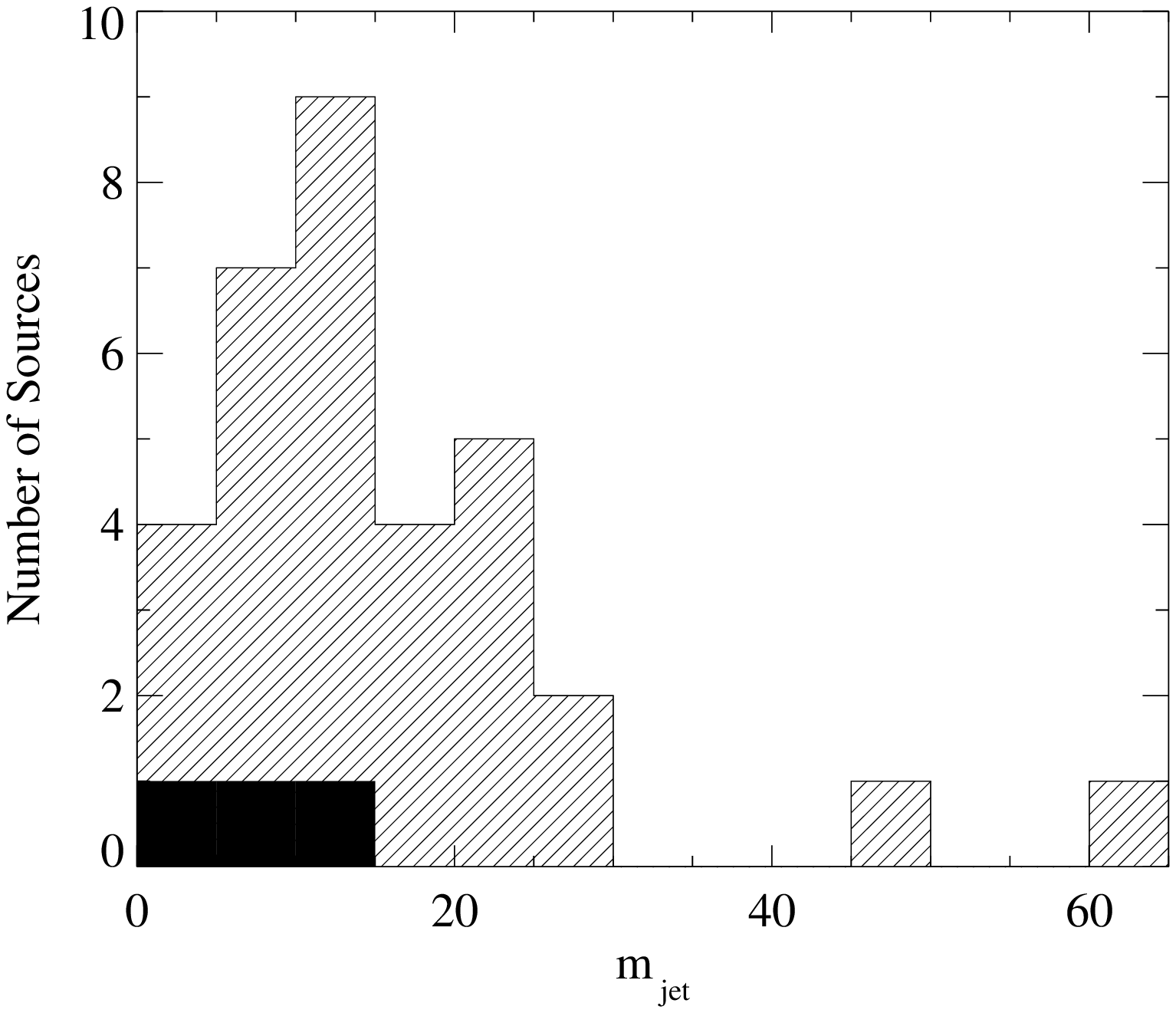}
\includegraphics[height=7.0cm]{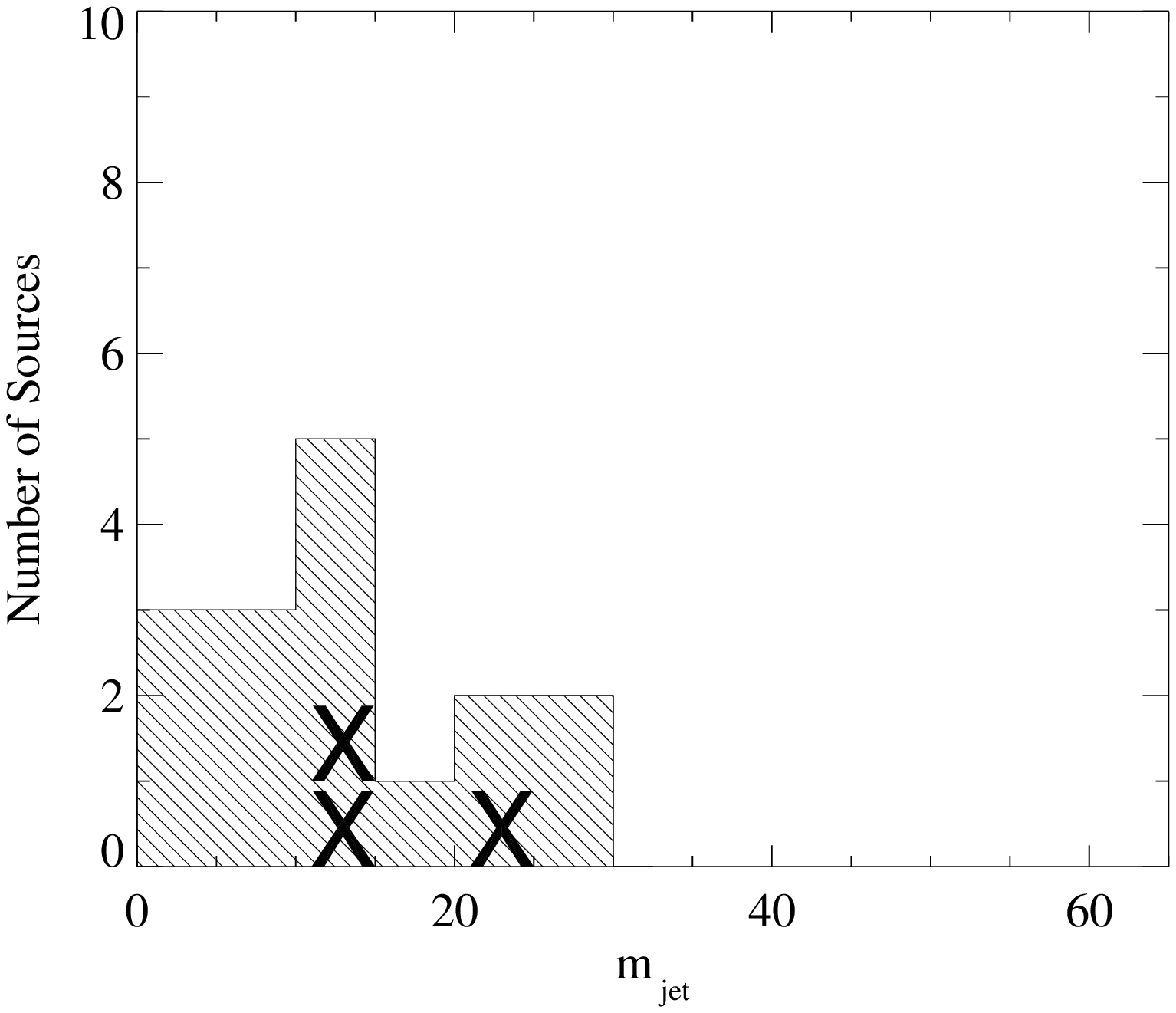}}
\caption{Distributions of inner-jet ($r < 10$ pc) fractional polarization 
for LBLs (left) and HBLs (right). The shaded regions mark the LBLs belonging
to our {\em HEAO-1}+RGB sample. X denotes a BL~Lac that is alternately
classified as an IBL by \citet{Nieppola06}.}
\label{fig:m_j}
\end{figure*}

Figure~\ref{fig:m_j} shows the distribution of the inner-jet fractional 
polarizations for LBLs and HBLs. The range of fractional polarizations in the 
inner jets (say, within 10 pc from the core), $m_j$, is the same for 
LBLs and HBLs, with the highest jet polarizations in both classes reaching 
tens of per cent, indicating the presence of highly aligned $B$ fields. 
The K--S test on the $m_j$ distributions for the HBLs and LBLs indicates
a $\sim$74\% probability that the two datasets come from the same 
parent population (Table~\ref{tab:KS}).

\begin{figure*}
\centerline{
\includegraphics[height=7.0cm]{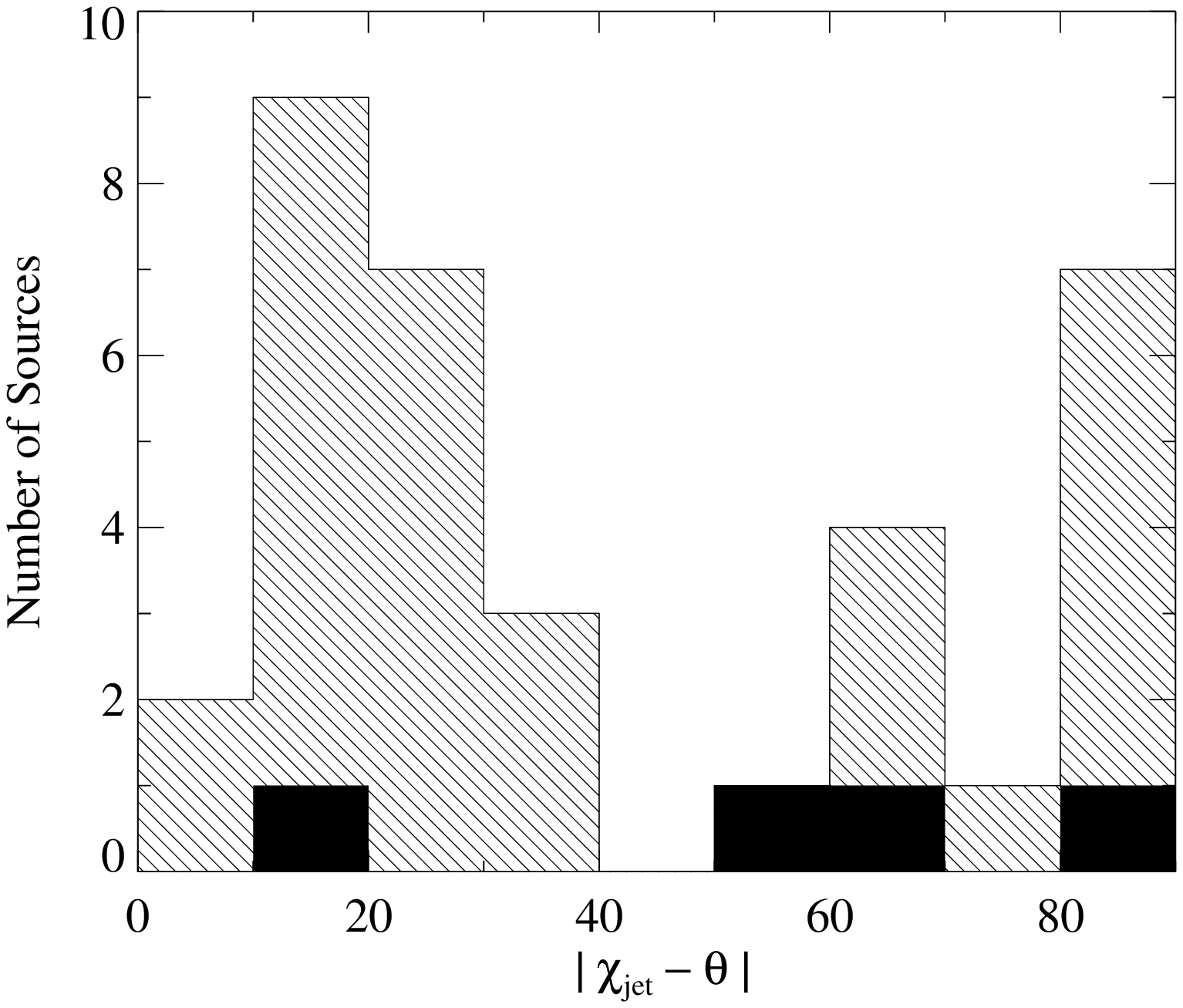}
\includegraphics[height=7.0cm]{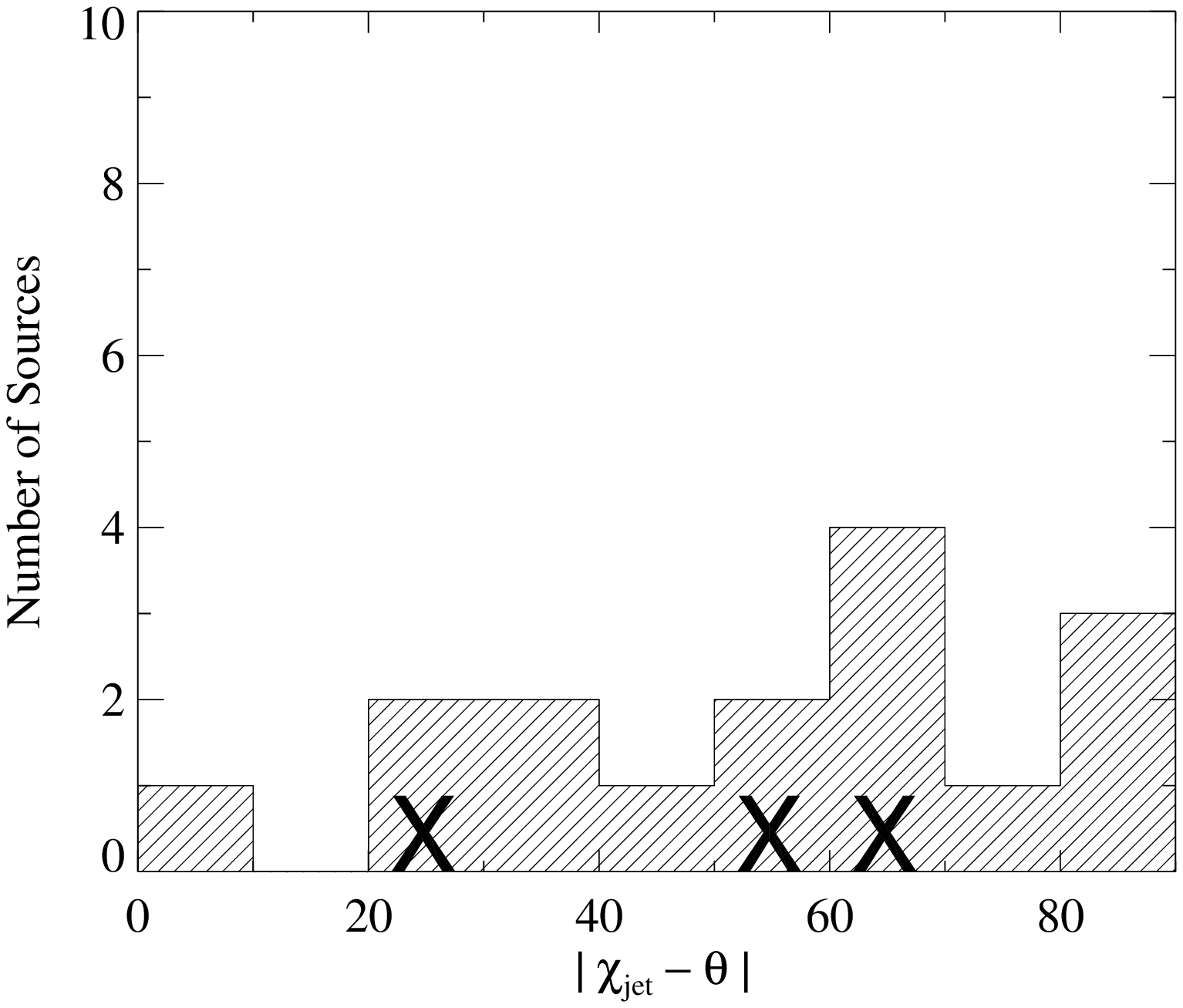}}
\caption{Distributions of offset between the inner-jet ($r < 10$ pc)
polarization angle
$\chi_j$  and the direction of the VLBI jet for LBLs (left) and HBLs
(right). The LBL distribution shows a predominance of values close 
to 0$^\circ$ and a secondary peak near 90$^\circ$. The HBL
distribution shows a large range of angles with a predominance of 
values larger than 60$^\circ$. The shaded regions mark the 
LBLs belonging to our {\em HEAO-1}+RGB sample. X denotes a BL~Lac that 
is alternately classified as an IBL by \citet{Nieppola06}.}
\label{fig:chi_j}
\end{figure*}

However, the jet polarization-angle orientation displays an intriguing
difference between HBLs and LBLs. While the jet EVPAs are predominantly 
aligned with the VLBI jet direction in LBLs, the EVPAs in HBLs show a 
tendency to be greater than 60$\degr$ with respect to the jet direction, 
that is, 
roughly perpendicular to the local jet direction (Fig.~\ref{fig:chi_j}).
The K--S test indicates a difference in the jet EVPAs between HBLs and
LBLs at the $\sim93$\% significance level. Assuming that the jet emission is 
optically thin, this result if true for HBLs, implies that
the HBLs show predominantly longitudinal jet $B$ fields, while  
the LBLs possess predominantly transverse jet $B$ fields. Clearly this 
bimodality in $B$ field structure needs to be tested with a larger sample of
HBLs. We note that the jet EVPAs of those HBLs which
have alternatively been classified as IBLs by \citet{Nieppola06}, occupy 
the middle of the EVPA range (between 20 - 70$\degr$).
This behaviour also warrants further investigation.

Some of the HBLs show evidence for a `spine-sheath' $B$ field structure, 
with the inner region of the jet having transverse $B$ fields and the edges 
having longitudinal $B$ fields. 
This type of $B$ field structure has been observed in other blazars
\citep{Attridge99,Giroletti04b} and it could result from interaction of the jet 
with the surrounding medium, or due to jet acceleration being a function
of the angular distance from the jet axis, producing a velocity
structure \citep{Ghisellini05}.
Alternatively, as \citet{GabuzdaMurrayCronin04} and \citet{Lyutikov05} have 
pointed out, this could be associated with the presence of a helical 
$B$ field associated with the jets of these objects. Such fields could
come about in a natural way due to the ``winding up'' of a seed field via the
combination of outflow and rotation of the central black-hole--accretion-disk 
system. Particularly good examples of a `spine-sheath' $B$ field structure
are 1227+255 (Fig.~\ref{fig:1227}) and 1727+502 (Fig.~\ref{fig:1727}). 

\subsection{Apparent Speeds}
We were able to derive tentative two-epoch apparent speeds for a number of
the objects in our sample from the model-fitting results in 
Table~\ref{modelBLL}, either on their own or combined with results from
the literature. We also derived firmer speeds for a smaller number of
objects based on three or more epochs, by combining our results with
other model fits in the literature.  The collected apparent
speeds for these and other HBLs from the literature are tabulated in
Table~\ref{tab:speed}, where Col.~(3) gives the jet component speeds, and
(4) the number of epochs used for estimating these speeds.

\begin{figure*}
\centerline{
\includegraphics[height=7.0cm]{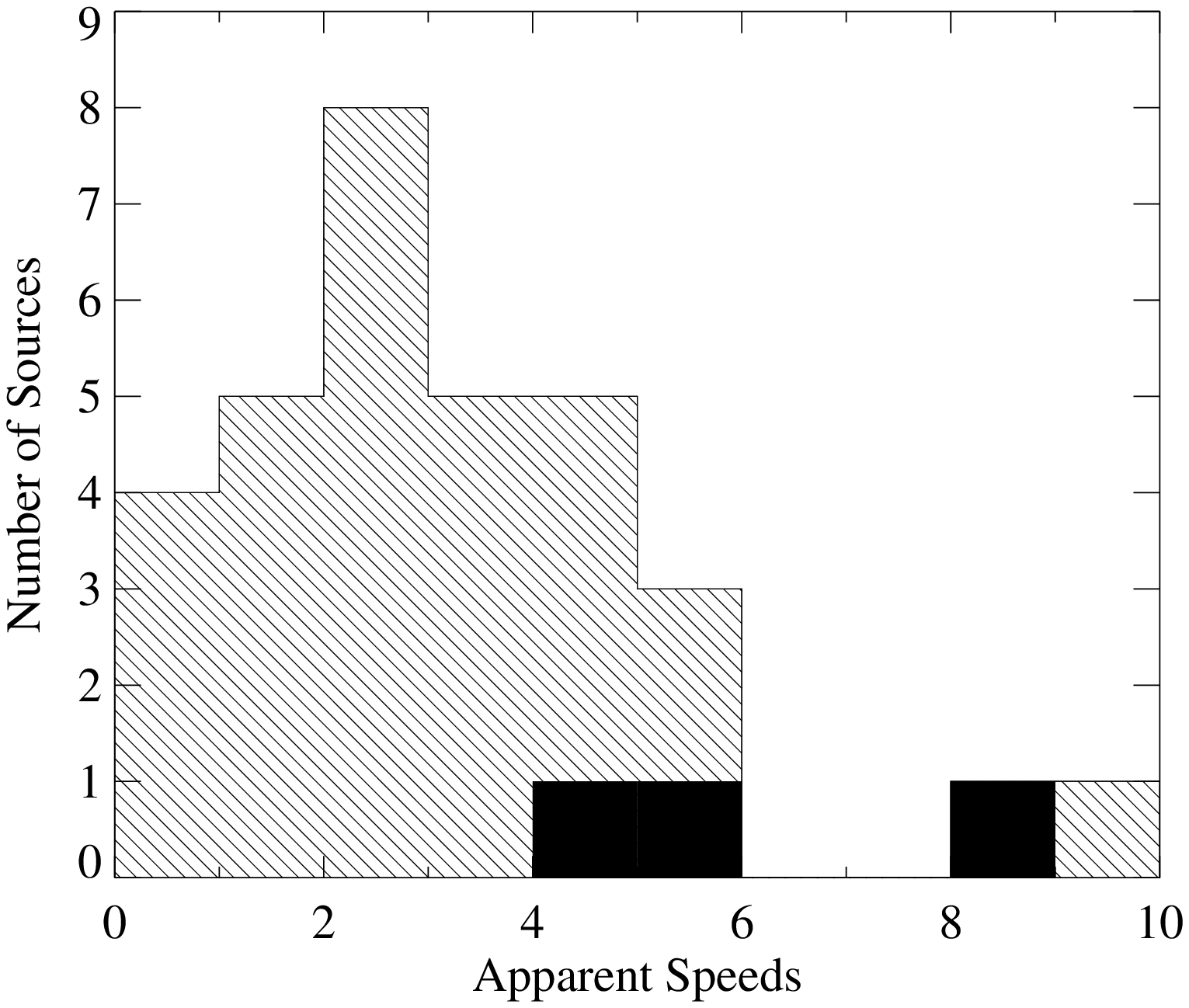}
\includegraphics[height=7.0cm]{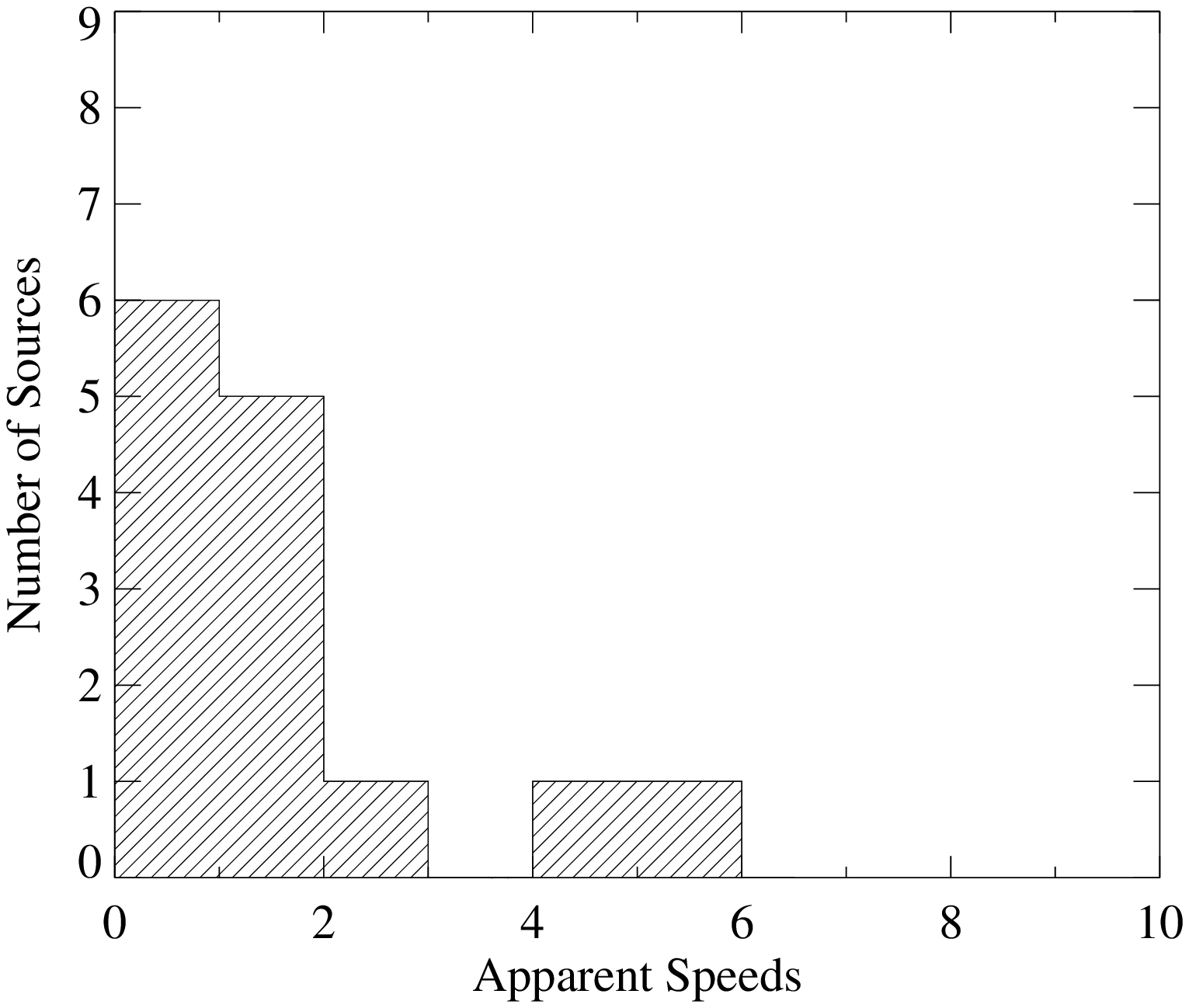}}
\caption{Distribution of apparent speeds of the jet components for the 1-Jy 
LBLs (left) and the HBLs (right). The apparent speeds in the LBLs show a peak 
between $2-3c$, somewhat higher than the peak of $1-2c$ seen in our HBLs. 
The shaded regions mark the LBLs belonging to our {\em HEAO-1}+RGB sample.} 
\label{fig:speeds}
\end{figure*}

LBLs typically show superluminal motions with apparent speeds in the 
range $1-5c$ \citep{Gabuzda00}. In Fig.~\ref{fig:speeds} we have 
plotted the apparent speeds of these, along with the LBL 0829+046 from the 
{\it HEAO-1} sample, shaded in black, which exhibits component speeds of 
$\approx5c$. Figure~\ref{fig:speeds} also shows the distribution of apparent speeds 
for our HBL sample. We find that $\beta_{app}$ is typically less than 2 for the HBLs. 
A two-sided K--S test indicates that the HBL and LBL speeds are different
at the 99.9\% significance level. This could suggest that either the HBL jets have 
lower Lorentz factors compared to LBLs, or they are oriented at larger angles to 
the line of sight.

\begin{table}
\begin{center}
\caption{Apparent speeds }
\begin{tabular}{ccccc}\hline\hline
Source    & Proper motion (mas/yr)  & $\beta_{app}$ & Number of Epochs & Ref\\ \hline
0414+009  & 0.13 &  1.81 & 2 & P  \\
0829+046  & 0.52, 0.54, 0.62, 0.88 & 4.95, 5.14, 5.90, 8.37 & 3&P     \\
1101+384  & 0.02, 0.03, 0.04, 0.05 &  0.03, 0.06, 0.09, 0.10  & 28 &  PE05\\
1133+704  & 0.04, 0.18, 0.78 & 0.11, 0.51, 2.23 & 4 & W    \\
1215+303  & 0.10, 0.16 & 0.75, 1.19&  2 & P   \\
1652+398  & 0.12, 0.25 & 0.26, 0.54& 12  & EP02, PE04   \\
1727+502  & 0.03, 0.16, 0.47, 1.53 & 0.10, 0.55, 1.63, 5.30 & 6, 6, 6, 5 & W \\
1741+196  & 0.04, 0.23 & 0.20, 1.17 & 5, 6 & W   \\
2155$-$304  & 0.117& 4.37 & 3 & PE04  \\
2344+514  & 0.044& 1.15 & 4 & PE04  \\
\hline
\label{tab:speed}
\end{tabular}
\end{center}
{Notes $-$ Reference for component speeds -- P = Present work, EP02 =
\citet{EdwardsPiner02}, PE04= \citet{Piner04}, PE05 = \citet{PinerEdwards05},
W= \citet{Wu07}.}
\end{table}

\begin{table}
\begin{center}
\caption{The Kolmogorov--Smirnov test statistic and probability 
for various parameters for HBLs and LBLs. }
\begin{tabular}{lllcllll}\hline\hline
Parameter & K--S statistic & Probability & Significantly Different?\\\hline
$L_{1.4}$ & 0.755 & 1.57E-05 & YES \\
$z$ & 0.655 & 0.00018 & YES \\
$m_{core}$& 0.655 & 0.003 & YES \\
$\beta_{app}$ & 0.589 & 0.001 & YES \\ 
$|\chi_{j} - \theta|$& 0.371 & 0.074 & YES?\\
$|\chi_{c} - \theta|$& 0.319 & 0.436 & NO \\
$m_{jet}$ & 0.198 & 0.738 & NO \\
\hline
\label{tab:KS}
\end{tabular}
\end{center}
\end{table}

\section{Discussion}

\subsection{Nature of the core polarization}
The HBL VLBI cores show lower fractional polarizations than LBL cores.
The lack of a correlation between core fractional polarization and
redshift at a high significance level however, suggests that
the higher $m_c$ values may be inherent to LBLs, and is not 
merely a resolution effect due to their higher average redshifts. 
There exists a possibility that the characteristic optical depths of 
the HBL cores are higher than those of the LBL cores, leading to lower
observed $m_c$ values for the HBLs. In this case, we might
expect to observe different behaviour in the distributions of $\chi_c$
for the HBLs and LBLs, which is observed in Fig.~\ref{fig:chi_c} albeit
for a very small number of sources. However it appears that the HBL cores 
could be intrinsically less luminous than LBLs, as also suggested by 
\citet{GiommiPadovani94} and \citet{PadovaniGiommi95}.
Figure~\ref{fig:power} and Table~\ref{tab:KS} demonstrate that HBLs are 
less luminous in their
total radio emission, while \citet{UrryPadovani95} have shown that
the same is true for their extended emission, supporting the 
picture wherein the radio cores in HBLs are intrinsically weaker than in LBLs.

\subsection{Jet polarization in HBLs and other Core-dominated AGNs}
\subsubsection{Differences in Viewing Angles ?}
Although larger viewing angles in HBLs were initially invoked to explain their
less extreme radio properties, this hypothesis was brought into question when
the HBL/LBL SEDs could not be reconciled simply with different viewing angles.
This debate is currently on $-$ while \citet{Rector03} have suggested larger
viewing angles for HBLs based on smaller pc-to-kpc-jet misalignments, 
\citet{Landt02} find no differences in viewing angles based on the 
Ca~H\&K break, an orientation-indicator. 

The lower apparent speeds observed in HBLs on their own could be
taken to support the different angle scenario, 
since lower observed apparent speeds could come about if the HBL jets
were viewed at larger angles to the line of sight (even if the intrinsic
speeds of the HBLs and LBLs were basically the same). However, we find that 
the other
differences that we observe between LBLs and HBLs cannot be reconciled
with this simple picture. This suggests that HBL jets are intrinsically
slower than LBLs. In this case, the HBL jets would have lower Doppler
factors, and so, on average, would be viewed at a larger range of
angles than the more Doppler-beamed LBL jets (but with this range being 
smaller than that for FRI radio galaxies). Thus, it is possible that the
main effect leading to the differences in the apparent speeds is a systematic
difference in intrinsic flow speed, but with the less important associated 
factor that the LBL jets are, on average, oriented at systematically 
smaller angles to line of sight, due to their increased beaming and 
subsequent prevalence in a flux-limited radio sample.

Note that \citet{Piner04} have suggested that the high-energy TeV emission 
from the TeV blazars (mostly HBLs), which requires high Lorentz factors, 
can only be reconciled with the slow jet speeds inferred from radio 
observations if their jets are decelerating fast on microarcsecond scales,
close to the central engine. This again supports the picture of 
intrinsically slower radio jets in HBLs compared to LBLs.

Several of our HBLs show evidence for `spine-sheath' $B$ field 
structures, with transverse $B$ field in the inner region of the jet and 
longitudinal $B$ field at the edges. Such structures have sometimes been
interpreted as reflecting the velocity structure in a jet with a 
faster `spine' and slower `sheath' (e.g. Attridge et al. 1999).
In this scenario, the more prominent longitudinal jet $B$ fields 
in HBLs could come about if their jets are oriented at relatively 
larger angles to the line of sight, where the emission from the 
slower-moving sheath field dominates, thus in principle supporting 
the ``different-angle scenario'' for HBLs and LBLs. 
However based on differences in the global properties of LBLs
and HBLs, it seems much more likely that, if the longitudinal $B$ field
at the jet edges is due to shear, this comes about for some other reason,
such as lower intrinsic speeds in HBL jets. In addition, as we consider
below, such `spine-sheath' $B$ field structures may have an entirely
different origin, having to do with the intrinsic $B$ field of the
jet.

\subsubsection{Spine-Shear Jet structure and Helical Magnetic Fields ?}
\citet{GabuzdaMurrayCronin04} and \citet{Lyutikov05} have
pointed out that `spine-sheath' structures of the sort observed for 1227+255
and 1727+502 could be observed if helical $B$ fields are associated with the 
jets of these objects. In this case, the dominant observed $B$ field
orientation (aligned with or perpendicular to the jet) would be determined
by the pitch angle of the helical $B$ field, and also to some
extent by the viewing angle. The pitch angle of the helical field might
plausibly be determined, for example, by the relationship between the 
speed of rotation of the central black-hole--accretion-disk system 
and the outflow speed of the jets. 
One way to understand the differences in the observed jet $B$ field 
structures of HBLs and LBLs is to suppose that this ratio tends to be 
lower for HBLs than for LBLs, resulting in smaller pitch angles for their 
jet $B$ fields. This could come about, for example, if the rotational 
speed were {\em lower} and the outflow speed {\em higher}
in HBLs than in LBLs. 

The simplest interpretation of our tentative finding that HBLs seem to 
have lower superluminal speeds than LBLs is that the HBL jets 
have lower bulk Lorentz factors than the LBL jets. If true, their
lower outflow speeds would tend to {\em increase} the pitch angles of the
associated helical $B$ fields (i.e., make their helical
fields more tightly wound), which would lead to a tendency for their
jet fields to be more, rather than less, dominated by a transverse field 
component. However, if the central rotational speeds of the HBLs are 
also lower than those in LBLs, the average ratio of the rotational to the
outflow speeds could end up being lower in HBLs than in LBLs, tending 
to give rise to a dominant longitudinal field component. 

In this picture, the characteristic jet Lorentz factors of HBLs would be 
systematically lower than those of LBLs and flat-spectrum radio quasars. 
It is interesting that the VLBI jets of FSRQs
also tend to show a predominance of longitudinal $B$ fields. One way
both the low-Lorentz-factor HBLs and the high-Lorentz-factor FSRQs 
could end up with predominantly longitudinal jet $B$ fields is if, 
in both cases, the ratio of the outflow to the central rotational speed 
is relatively high --- in the HBLs due to relatively low rotational 
speeds, and in the quasars due to relatively high outflow speeds. 

Indeed, the observed superluminal motions in quasar jets are, on average, 
somewhat higher than those in LBL jets \citep{Gabuzda00,Kellermann04}; 
the simplest 
interpretation of this systematic difference is that the typical
jet Lorentz factors are higher in quasars than LBLs. Lower apparent speeds
in LBLs could also come about if their jets were viewed at somewhat larger
(or smaller) angles to the line of sight, but the former possibility is at
odds with the high polarization and variability of LBLs, and the latter
possibility would imply that LBLs should have higher Doppler factors than
quasars, which is not the case. If HBLs display lower superluminal speeds 
than LBLs, and this reflects lower typical jet Lorentz factors in HBLs 
compared to LBLs, we find that FSRQs, LBLs and HBLs form a sequence of 
decreasing average jet Lorentz 
factor. Note that the order in this sequence is the same as
that for the sequence in emission-line luminosity, from highest in FSRQs 
to lowest in HBLs \citep{Ghisellini97}, and also for the sequence of the 
synchrotron peak frequencies, from lowest for FSRQs to highest for HBLs. 

If a jet does have a helical $B$ field, this should 
give rise to a systematic gradient in the observed Faraday rotation 
measure across the jet, due to the systematic change in the line-of-sight
$B$ field component \citep{Blandford93}, as has been observed for several
LBLs \citep{GabuzdaMurrayCronin04}. Multi-frequency polarization VLBA data
for the {\em HEAO-1}+RGB BL Lac objects considered here are currently being
reduced; the detection of systematic rotation measure gradients transverse 
to the VLBI jets of any of these sources would appreciably strengthen the case 
that they are associated with helical $B$ fields. 

\citet{Meier97} and \citet{Meier99} have demonstrated through numerical
simulations that there may be a connection between the output radio power and the 
spin rate of the central black hole. Although the original idea was to explain the 
FRI and FRII classes, a similar idea could apply for the BL~Lac subclasses. Objects
with lower radio powers (i.e., the HBLs) could have black holes spinning at lower 
rates, which produce slower jets, while objects with higher radio powers 
(i.e., the LBLs) could have black holes spinning at higher rates, which produce 
relatively faster jets. In this picture, the unresolved bases of the jets, 
$i.e.,$ the 
radio ``cores'', should be weaker in the HBLs, in agreement with our
observations, while the radio cores of LBLs should be more dominant.
We have already discussed above how differences in the black-hole spin 
rates of HBLs and LBLs could lead to the observed differences in 
VLBI polarization structure, if the jets have helical magnetic fields.
In addition, there is the secondary effect that
the less Doppler-beamed HBL jets would be oriented, on average, at 
somewhat larger angles to the line of sight than the 
more beamed LBLs (but always within a smaller range than for their unbeamed 
counterparts, the FRI radio galaxies). Thus, through all the 
observational manifestations and consequences that could plausibly follow
from a difference in characteristic black-hole spin rates, this  
one most fundamental difference may essentially be able to  
explain the observed properties of LBLs and HBLs. We will be exploring 
this scenario as a means of understanding the $B$ field
geometries observed in quasar subclasses as well \citep{Lister01}
in a future paper.

\subsubsection{Misclassified Quasars in the BL~Lac populations?}
Some HBLs and LBLs have been found to have FRII radio powers and morphologies 
\citep{Kollgaard92,Murphy93}, and it is possible that some 
subset of the HBLs and LBLs are actually misclassified radio quasars.
However, there is no tendency for the HBLs that have longitudinal jet
$B$ fields also to have the lowest core polarizations, arguing against the
idea that a large fraction of the HBLs are misclassified quasars. In 
addition, the observed superluminal speeds in quasars are clearly higher than 
the tentative speeds available for HBLs thus far.

\subsubsection{Comparisons of BL~Lac subclasses with similar $B$ field structures}
\citet{Gabuzda00} have demonstrated that the pc-scale jet EVPAs in LBLs 
show a bimodal distribution $-$ while the majority of LBLs exhibit
transverse jet $B$ fields, there is a subset of LBLs that exhibit a longitudinal 
$B$ field morphology (see Fig.~\ref{fig:chi_j}). We compared the properties of 
the HBLs with the
subset of LBLs that show longitudinal $B$ fields similar to the HBLs.
We found that like the majority of LBLs, this subset of LBLs still 
show all the differences in their basic properties
which are observed in the two subclasses as a whole, i.e., the LBL subset still
have systematically higher redshifts, higher core fractional polarizations, 
greater component speeds, and no difference in the distribution of the jet
fractional polarizations or core EVPAs compared to HBLs.
This strongly supports the picture of intrinsic differences in the LBL
and HBL classes.

\section{Conclusions}

\hspace*{2em}1.~~We have observed eighteen BL~Lacs belonging to the {\it HEAO-1} and 
RGB samples with VLBI, and detected parsec-scale polarization 
in all but three of them. After dividing each of
our objects into the HBL or LBL classes and considering the known
results on LBLs in the literature, we have carried out a comparison of
the relevant properties of the two BL~Lac classes.

2.~~We find that the total intensity pc-scale images reveal a 
core-jet morphology, similar to that observed in LBLs.

3.~~The VLBP observations of HBLs and LBLs reveal differences in their
core fractional polarization, similar to the trend observed on
kpc-scales -- the HBLs have a lower core fractional polarization
compared to the LBLs.  High degrees of core polarization were observed 
for the two RGB sources 0749+540 ($\simeq 10\%$) and 0925+504 
($\simeq 14\%$), but both of these are, in fact, LBLs.  

4.~~The jet fractional polarizations in HBLs do not differ 
systematically from those in LBLs. In both cases, jet degrees of polarization
of tens of per cent can sometimes be observed, indicating the presence of
well-ordered $B$ fields.

5.~~The relative VLBI core polarization angles do not show any systematic
differences between HBLs and LBLs. 

6.~~The most intriguing difference between the HBLs and the LBLs is 
the orientation of the predominant $B$ fields in their 
pc-scale jets -- the $B$ fields in HBL jets tend to be aligned with 
the local jet direction, while the $B$ fields of LBLs 
tend to be perpendicular to the local jet direction.
This result needs to re-examined with a larger sample of HBLs.

7.~~Some of the HBLs display evidence for jet magnetic field structures with
transverse $B$ field in the inner region of the jet and longitudinal $B$
field at the edges, i.e, a `spine-sheath' structure. Although such $B$-field
structures have been taken to indicate a fast spine+slow sheath velocity
structure, they can also come about in a natural, simpler way if the jet has
a helical $B$ field (without any requirement for a two-layer velocity
structure).

8.~~We find tentative evidence that the observed component speeds in HBLs 
are lower than the typical apparent speeds  observed in LBLs ($\simeq 1-5c$)
and FSRQs ($\simeq 5-10c$).  This result suggests that either the HBL jets 
have lower Lorentz factors than LBLs and FSRQs, or that their jets are oriented 
at larger angles to the line of sight. Since the collected properties of
HBLs provide no clear evidence that their jets lie at systematically 
larger angles to the line of sight than do LBL jets, this suggests that
the outflow speeds are intrinsically lower in HBLs, than in LBLs.

9.~~One way to understand our collected results in the context of other
previous results in the literature is if HBLs, LBLs and FSRQs form a sequence 
of increasing average jet Lorentz factor --- the same order as for the 
sequence of the synchrotron peaks of these objects, which proceed from
highest peak frequency for HBLs to lowest peak frequency for FSRQs, although
the physical connection between these two sequences is not clear.

10.~~If the jets of all three classes of compact AGN characteristically 
have helical $B$ fields due to the ``winding up'' of a seed field via the 
combination of outflow and rotation of the central black-hole--accretion-disk 
system, then the dominant jet $B$ fields observed in these three 
classes of compact AGNs $-$ longitudinal in HBLs and FSRQs and 
transverse in LBLs $-$ 
could come about due to the ratios of their characteristic 
rotational and outflow velocities. If HBLs, LBLs and FSRQs do form a 
sequence of increasing Lorentz factor, the low outflow speeds in HBLs 
would require relatively slow rotation ($i.e.,$ smaller spin rates) of their central 
black-hole--accretion-disk systems. 

\section*{Acknowledgments}
We would like to thank the referee, Eric Perlman, for a careful assessment of our
work, which has greatly improved the paper.
PK is grateful to the Joint Institute for VLBI in Europe (JIVE) for 
providing her with a summer studentship in 2001. We would like to acknowledge 
Ron Kollgaard's contribution in initiating the VLBP project.
This research has made use of the NASA/IPAC Extragalactic Database (NED) 
which is operated by the Jet Propulsion Laboratory, California Institute of 
Technology, under contract with the National Aeronautics and Space 
Administration. 

\bibliographystyle{mn2e}
\bibliography{ms}

\bsp

\label{lastpage}

\end{document}